\documentclass[preprint,12pt]{elsarticle}

\usepackage[utf8]{inputenc}
\usepackage{acronym}
\usepackage{graphicx}
\usepackage[tight,footnotesize]{subfigure}
\usepackage{color}
\usepackage{array,multirow,hhline}
\usepackage{xspace}
\usepackage[cmex10]{amsmath}
\usepackage{lscape}
\usepackage{epstopdf}
\usepackage{float}

\usepackage{array}
\newcolumntype{C}[1]{>{\centering\let\newline\\\arraybackslash\hspace{0pt}}m{#1}
}

\usepackage[a4paper]{geometry}

\acrodef{ABPS}{Always Best Packet Switching}
\acrodef{AP}{Access Point}
\acrodef{CN}{Correspondent Node}
\acrodef{MN}{Mobile Node}
\acrodef{NIC}{Network Interface Card}
\acrodef{QoS}{Quality of Service}
\acrodef{SIP}{Session Initiation Protocol}
\acrodef{TED}{Transmission Error Detector}
\acrodef{HA}{Home Agent}
\acrodef{CoA}{Care of Address}
\acrodef{VoIP}{Voice over IP}
\acrodef{NAT}{Network Address Translation}
\acrodef{GPL}{General Public License}
\acrodef{API}{Application Programming Interface}
\acrodef{STUN}{Session Traversal Utilities for NATs}
\acrodef{TURN}{Traversal Using Relays around NAT}
\acrodef{ICE}{Interactive Connectivity Establishment}

\journal{Computer Networks}

\begin{document}

\begin{frontmatter}

\title{A Survey on Handover Management in Mobility Architectures}

\author{Stefano Ferretti}
\ead{s.ferretti@unibo.it}

\author{Vittorio Ghini}
\ead{vittorio.ghini@unibo.it}

\author{Fabio Panzieri}
\ead{fabio.panzieri@unibo.it}

\address{Department of Computer Science and Engineering, University of Bologna\\ Mura Anteo 
Zamboni 7, 40127 Bologna, Italy}

\begin{abstract}
This work presents a comprehensive and structured taxonomy of  available techniques for managing the handover process in mobility architectures. 
Representative works from the existing literature have been divided into appropriate categories, based on their ability to support horizontal handovers, vertical handovers and 
multihoming. We describe approaches designed to work on the current Internet (i.e.~IPv4-based 
networks), as well as those that have been devised for the ``future'' Internet 
(e.g.~IPv6-based networks and extensions). 
Quantitative measures and qualitative indicators are 
also presented and used to evaluate and compare the examined approaches.
This critical review provides some valuable guidelines and suggestions for designing and developing mobility architectures,  
including some practical expedients (e.g. those required in the current Internet environment), aimed to cope with the presence of NAT/firewalls and to provide support to legacy systems and several communication protocols working at the application layer.
\end{abstract}


\end{frontmatter}

\section{Introduction}

Over the last few years, several architectural solutions have been proposed to support users that connect to the Internet through a \ac{MN}. 
The main objective is to provide seamless communications, 
i.e.~ensuring that if a \ac{MN} changes its point of attachment to the Internet, while in movement, no communication interruptions are perceived at the application 
level, and if such interruptions occur, they do not significantly degrade the 
\ac{QoS} delivered at the application level.  
While throughput remains a major goal of system design,
the main concern of mobility architectures is how to best manage situations where a MN changes network. This event is 
currently referred to as \textit{handover} (or \textit{handoff}).  

By default, current operating systems installed on smartphones adopt the following strategy for data transmission: one \ac{NIC} at a time is configured and employed 
to send data. If a WiFi network is available, the terminal switches to WiFi; otherwise a cellular network is 
utilized, if the latter is available too. 
During the handover, communications are interrupted. While the widespread use
of current smartphones confirms that in general such a simple approach may be a viable solution, in some cases this strategy has some severe limitations.
Just as an example (which is actually a true story), let us consider the case of an employee working in an institution/company composed of several buildings, all covered by a WiFi network (e.g.~a researcher in a university campus). 
Suppose that the researcher is a commuter and, just before leaving for going home, he/she receives an important \ac{VoIP} 
phone call. Since he needs to leave to take the last train home, he decides to answer the 
call using his/her mobile phone; today, there are plenty of smartphone apps that 
offer very efficient VoIP services.
At that moment, the device is connected through WiFi, but when he gets out of the 
building, the WiFi signal is lost and the smartphone automatically switches to 4G without any handover management at the application level, 
thus experiencing a first communication interruption.  
While moving, he passes through other buildings (hence, within their WiFi 
coverage); as a consequence, the smartphone switches back to WiFi 
(i.e.~a second communication interruption occurs), and then back to 4G 
(i.e.~yet another communication interruption) and so on. 
One might suggest that the employee should turn off the WiFi NIC before leaving, thus using the cellular NIC only; yet, a full 3G/4G coverage may not be available in all the various buildings he goes through.
Indeed, when moving, there are cases where one needs to change NIC without breaking the communication at the application layer.

Handovers may actually require a change of the \ac{AP} (cell or WiFi AP) due to the MN 
movement, thus causing a reconfiguration at the datalink layer of the \ac{NIC} 
in use (horizontal handover). However, the MN can also change the network technology, 
switching from one \ac{NIC} to another, which causes a reconfiguration at 
higher network layers (vertical handover). 
Changing network means that the IP address associated to the \ac{MN} changes as well; 
this has repercussions on the application layer, since in the current Internet, 
the IP address of a node usually plays the twofold role of MN locator and MN 
identifier. 
This ``change of identity'' causes a service interruption that requires more 
time than a simple network reconfiguration at the operating system 
level. 

Various proposals have been put forward to deal with 
this problem. Some approaches described in these proposals suggest a decoupling of the 
node identifier from its address (locator), e.g.~GLI-Split \cite{gli-split}, 
HIP \cite{BokorZNJ09,rfc4423}, Hi3 \cite{GurtovKLN08}, ILNP 
\cite{rfc6740}, LISP \cite{lisp}, MILSA \cite{Pan:2010}, NIIA 
\cite{niia,Schutz:2010}, RANGI \cite{rangi}. These approaches usually comply 
with future Internet visions, requiring some radical changes in the network 
architecture.
Other approaches address the above mentioned ``change of identity'' issue by exploiting (and enhancing) protocols of the 
current Internet stack, e.g.~ABPS \cite{GhiniJSS}, DCCP \cite{rfc4340}, 
SIP-IAPP \cite{802f02,WuYH07},
I-TCP \cite{itcp},
MMUSE \cite{SalsanoPMNV08},
MPTCP \cite{mptcp,Paasch:2012}, 
m-SCTP \cite{Budzisz:2012}, 
MSOCKS \cite{msocks}, 
TCP-migrate \cite{snoeren2001reconsidering}.
These latter approaches can be classified based on their ability to support the use of a single \ac{NIC} at a time, or the 
(possibly concurrent) use of multiple \acp{NIC}.  
They may work at various levels in the network 
stack, or even use a cross-layer strategy employing different functionalities at 
different levels. 

The plethora of available proposals reveals that there are many technical issues concerned with the main problem of mobility management, as well as different technologies that need to be supported, and several alternative ways of solving these issues and using the technologies available. 
There are solutions that may work in principle, but cannot be deployed in practice because, for example, 
i) the protocols they use do not comply with the current Internet implementation, 
ii) they cannot deal with the situation in which a node is behind a \ac{NAT}/firewall, and
iii) they do not take into consideration the fact that some important 
existing applications do not respect the Internet protocol stack stratification.
Thus, there is a significant need to identify and state the main issues making up the whole problem, and to classify the possible approaches for mobility management.
This paper illustrates these main issues, as well as experiences and 
lessons learned from systems and proposals available in the literature, and eventually
provides a critical discussion that might help practitioners in devising a 
holistic solution for mobility management support.

In the rest of the paper, we give some background information on host mobility management services,  review the main architectural solutions proposed in the literature, and come up with a classification that arranges solutions according to their design principles. We also discuss aspects that have an impact on their deployment in real scenarios and limit their applicability. 
It is worth mentioning that, while this paper was being written, new studies have been published on the same topic, e.g.~\cite{survey}. 
However, these works mainly focus on the aforementioned future generation Internet and on locator-identifier separation mechanisms. Instead, our approach emphasizes multihoming, mobility, the possibility of easily switching from one NIC to another, the compatibility with the existing Internet and problems strictly concerned with the limitations of the current applications and the architectural solutions employed (e.g.,~presence of NATs, firewalls, violations of the protocol stratification).
Moreover, in this paper we focus on host centric 
networks, i.e.~the traditional host-based conversation model that is exploited 
in the current Internet. Thus, we do not consider neither information-centric networks 
\cite{icn} nor user-centric networks \cite{tuncer,Wang:2011}. 

The main contributions of this work are the following.
\begin{enumerate}
  \item We provide a comparative overview of the main architectural solutions 
for mobility support in wireless networks.  
  All the considered systems are classified based on their main 
characteristics, the features they offer, and the level of network protocol stack they operate.
  \item We provide some valuable guidelines for developing mobility 
architectures in the current Internet, summarised as follows. Firstly, proxies used in many applications 
(VoIP, \ac{SIP}-based applications, and optionally HTTP-based applications 
\cite{FerrettiG09}) should be upgraded/extended to cope with mobility issues. 
Secondly, NATs and firewalls have to be handled carefully. Thirdly, 
multihoming solutions should take into account that many widespread 
applications and related protocols (e.g.~\ac{SIP}-based ones) do violate the layering principles of the protocol stack. Solutions to this problem require the use of an 
external proxy and/or the modification of application messages.
\item We describe the main quantitative measures and qualitative indicators for evaluating mobility 
architecture, and classify all the presented approaches accordingly.
\end{enumerate}
The remainder of this paper is structured as follows.
Section \ref{sec:background} provides the background information and the basic 
definitions related to this topic.
Section \ref{sec:singleNIC} presents the existing architectural solutions 
working with single \acp{NIC}, while Section \ref{sec:multihoming} discusses 
solutions that exploit multiple \acp{NIC}. 
Section \ref{sec:comparison} gives a qualitative comparison of the host 
mobility architectures discussed in the paper. 
Finally, Section \ref{sec:conc} provides some concluding remarks and the main guidelines 
for developing mobility architectures.

\section{Main Definitions and Concepts}
\label{sec:background}

The aim of a host mobility architecture is to ensure that a MN can move seamlessly across 
different access networks, without any interruptions of the active network services. Before going 
into the details of such architectures, in this section we need to introduce some concepts and terminology that will be used throughout the whole paper. Then, we will describe how handover is executed when a MN changes its AP within the same Internet Service Provider (ISP). We will also discuss how MNs are identified and localized in a network. We will clarify the notion of proxies and relays, which are useful distributed entities for supporting mobility. After that, we will propose a classification of handover management schemes to support mobility. From the discussion provided, it will be clear that dealing with NATs and firewalls (for systems working on the current Internet) is absolutely crucial and that several implications arise from the use of session and application layer protocols that cannot be ignored.

\subsection{Definitions}

\subsubsection*{Handover}
Handover (or handoff) is the process of transferring a network communication 
when a mobile terminal changes its connection point to the access network 
(called ``point of attachment''). 
\subsubsection*{Seamless Handover}
A seamless handover occurs when the handover is performed with no user 
perceivable interruptions, hence guaranteeing that the user communication 
session remains active. 
\subsubsection*{Horizontal Handover}
\begin{figure}[th]
   \centering
   \includegraphics[width=.6\linewidth]{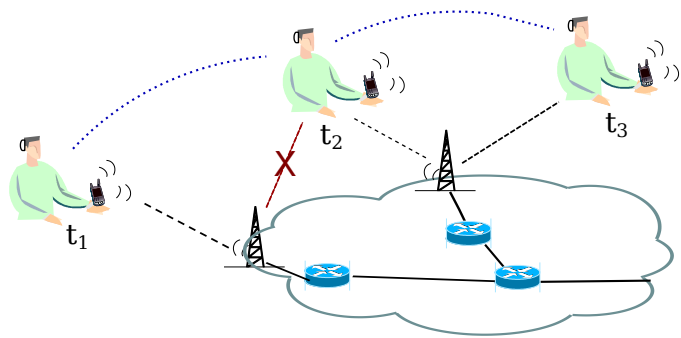}
\caption{Horizontal Handover: a MN changes access network using the same NIC, 
while moving}
   \label{fig:h_handover}
\end{figure}

A horizontal handover takes place between points of attachment supporting the 
same datalink-layer network technology (Figure \ref{fig:h_handover}). A user 
moving from one UMTS cell to another UMTS cell is an example of a horizontal 
handover.

It is worth mentioning that there are different types of horizontal handovers. 
For example, in cellular and WiMAX networks, the horizontal handover can be 
\emph{hard} or \emph{soft}.
\subsubsection*{Hard Handover}
A hard handover is (a horizontal handover) designed to first break off the 
initial connection with an 
AP (base station), before switching to another one. Thus, the 
MN communicates with one AP at a time only. Connection with 
the old AP is broken before the new connection is established. 
A hard handover is also referred to as ``break before make'' handover, and it 
should result in a non-perceptible and short interruption 
(i.e.~even if it is a hard handover, it should be seamless).
\subsubsection*{Soft Handover}
With a soft handover, the connection between the MN and its AP (base station) is retained until a connection is established with another AP. This mechanism is also known as “make before break”. In that way, the MN may be connected with two (or even more) APs at a given moment in time.
\subsubsection*{Vertical Handover}
\begin{figure}[th]
   \centering
   \includegraphics[width=.6\linewidth]{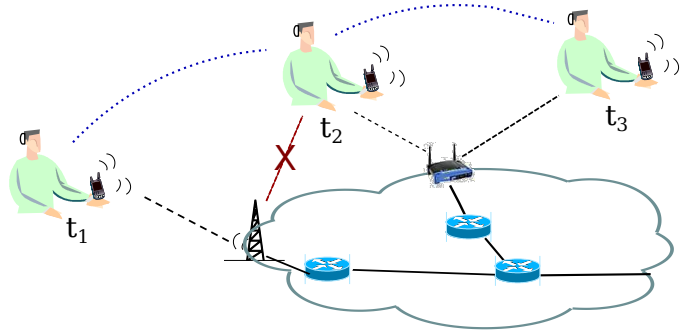}
\caption{Vertical Handover: a MN changes access network and the NIC used, while 
moving}
   \label{fig:v_handover}
\end{figure}

Vertical handover occurs between points of attachment supporting different 
datalink-layer network technologies (Figure \ref{fig:v_handover}). The switch 
from WiFi to a cellular network, and vice versa, are examples of vertical handovers.
\subsubsection*{Multihoming}
\begin{figure}[th]
   \centering
   \includegraphics[width=.8\linewidth]{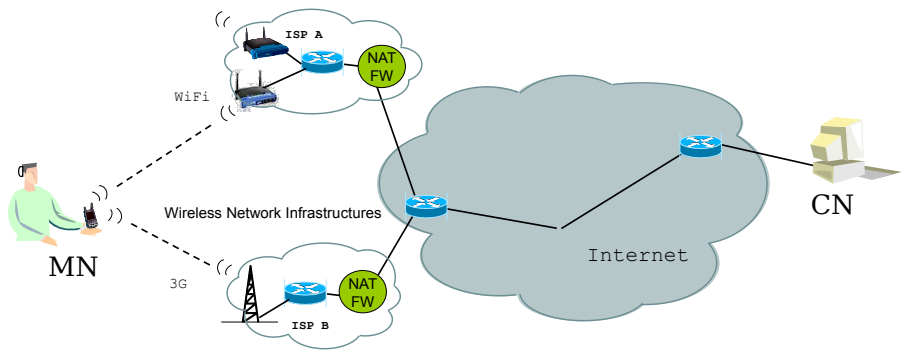}
\caption{Multihoming: a MN has different active NICs available, ready to be 
used}
   \label{fig:mhoming}
\end{figure}

Today, a MN is equipped with multiple NICs (and corresponding IP addresses). 
Multihoming refers to the possibility that a MN is connected simultaneously to 
more than one network (Figure \ref{fig:mhoming}). 


Various sophisticated approaches exploit multihoming as a basic solution to increase 
reliability and offer seamless communications. Ideally, a 
seamless mobility architecture is responsible for: 
\begin{enumerate}
  \item identifying each given MN univocally; 
  \item allowing each MN to communicate with its \ac{CN}, (for the purposes 
of this discussion, a CN might be a fixed or mobile node communicating with 
the MN); 
  \item monitoring the QoS provided by different networks, predicting the need of a handover and selecting a new preferred NIC (and the related access point);
  \item performing the handover seamlessly, i.e. ensuring the continuity of the 
communications without any perceivable interruptions for the end users. 
\end{enumerate}

\subsection{Intra-ISP Handovers}
This work focuses on systems that deal with the change of network domain, while moving. Thus, we examine mechanisms that allow the connectivity to be maintained when a MN changes its Internet Service Provider (ISP). 
Before going into the details of such systems, it is important to understand how handovers are handled in current cellular networks and WLANs, when a MN changes its AP within the same ISP. Reconfigurations take place at the physical and data link levels only, while the IP address of the MN remains unchanged.

These mobility management schemes involve the same NIC and the same 
carrier/network, i.e. they are horizontal handovers. 
It is worth mentioning, however, that some works refer to the change between 3G 
and 4G as a vertical handover, since, even though the MN employs the same NIC to 
communicate, there is a change in the type of connectivity used to access a
supporting infrastructure  \cite{Tu:2014}.

\subsubsection{Cellular Networks}
A cellular network architecture consists of a set of base stations and a core network. The base stations are the APs that provide radio access to MNs, whereas the core network connects them to the wired Internet or the public telephone network. The core network has a very different structure depending on the type of cellular network. For example, 3G networks such as UMTS support both circuit switched services, used to carry traditional voice communication, and packet switched services for data traffic. 
The new 4G Long-Term Evolution (LTE) networks, on the other hand, have been designed to offer packet switched data services only. Voice services should be delivered as data flows over LTE (Voice over LTE -- VoLTE). 
However, due to the high deployment cost and complexity of making such a transition, most 4G operators currently adopt Circuit-Switched Fallback (CSFB), which switches 4G users to legacy 3G and accesses the circuit switched voice service in 3G 
\cite{Tu:2014}.

A detailed description of the cellular communication architecture is out of 
the scope of this paper. In a few words, a handover performed within the same 
ISP consists of an interaction between the MN and the 
core network to select the base station employed to transmit and receive data. 
In UMTS, for instance, handovers are managed through the Universal 
Terrestrial Radio Access Network (UTRAN), while LTE uses an Evolved UTRAN 
(EUTRAN), which simplifies the handover management, thus ensuring lower 
latencies for handovers and connection setup.

A different procedure is required when the MN switches between 3G and 4G. 
This procedure is divided into two parts, i.e.~preparation and 
execution. 
During the handover preparation, resources are reserved in the target network.
In the execution phase, the MN is handed over to the target network from the 
source network. Thus, only the network path needs to be switched, reducing 
handover processing time.
This is done to avoid packet losses, even between radio access systems such as 
LTE and 3G, which cannot be used simultaneously.

\subsubsection{WLANs}
Here we focus on the horizontal handover performed when a MN switches from an 
AP of a given WLAN to another AP of the same network, i.e.~between APs using 
the same Service Set Identifier (SSID).
The IEEE 802.11 does not specify how to handover from one AP to another one. 
However, the IEEE released a recommendation called Inter-Access Point Protocol 
(IAPP) \cite{802f02}, according to which the handover between different APs follows four steps: i)
when a MN finds an AP with a signal which is better than the one it is 
currently attached to, it breaks the connection with the current AP and sends 
a reconnection request to the target AP; 
ii) the target AP establishes a new connection with the MN and sends a 
notification to the current AP; iii) on reception of this notification, the 
current AP transfers the MN information to the target AP; iv) finally, the MN 
switches to the target AP.
This is done transparently to the application, and no changes 
are required at the end points (however, APs must run an implementation of the 
IAPP protocol).
It is worth mentioning that this protocol 
has been extended to support handovers in different subnets \cite{TMC.2005.49} 
and between different WLANs \cite{WuYH07}; the latter solution relies on the use of the \ac{SIP} protocol.

There are alternative (and in some cases proprietary) solutions to this 
approach, e.g.~the Fast Secure Roaming offered by Cisco, and other scientific 
proposals such as \cite{ghini_wdays10,pack,ramani}.
Amongst others, it is worth mentioning the IEEE 802.11r-2008, also called 
Fast Transition, which is an IEEE standard for performing a soft horizontal 
handover \cite{802.11r}. It is an amendment of 
the IEEE 802.11 for fast roaming, where the initial handshake with the new AP 
is done before the MN roams to the target AP. IEEE 802.11r redefines the 
security key negotiation protocol executed when a MN decides to connect with 
a new AP. In essence, both the negotiation and the request for wireless 
resources occur in parallel. This will remove much of the handshaking overhead 
while roaming, thus reducing the handover time between APs while providing 
security and QoS.

\subsection{Node Localization}

In IP-based communications, the MN’s IP address plays the twofold role of MN 
identifier and locator, as it distinguishes the MN uniquely and identifies 
its position in the network. 
When a MN joins a network, it is assigned an IP 
address that is valid within that network only. When that MN moves to a 
different network, it acquires a new IP address, thus losing its identity. 
Hence, it needs to inform CNs of its new identity and location. In general, it 
is not possible to exclude that a CN will move while communicating with the 
MN; for that reason, a mobility management architecture must ensure that two hosts can 
communicate even when they both move and change their addresses simultaneously.

The above problem arises because the IP protocol and its related addressing scheme 
were originally developed for wired (non-mobile) networks. A number of approaches address this problem by adding and/or modifying services at the network level.
In particular, different solutions proposed in the literature attempt to overcome 
the above problem by separating the node identifier from the node locator 
\cite{lisp,Pan:2010,Atkinson:2010,Gladisch:2014,Ramirez:2014,rfc5533,Tuncer:2013}.
The mobile node will therefore have a single identifier that may change over time, regardless of its location. 

To summarize, all mobility management architectures adopt some mechanism 
that: 
\begin{enumerate}
 \item defines a \textbf{MN identifier} uniquely, no matter what the location 
of the MN is, and 
 \item provides a \textbf{localization service} that maintains a mapping 
between the MN identifier and the current MN location, even when the MN moves. 
\end{enumerate}
The localization service is based on a location registry, a service that is 
assumed to always be available. When a MN changes its IP address, it 
passes the information to the registry, that is thus informed of the change
(\textbf{registration phase}). 
When a CN wants to initiate a new communication with the MN, or when it wants 
to continue a communication with the MN that just changed its address, the CN 
starts a \textbf{lookup phase}, asking the location registry for the MN’s 
current address, and then uses the obtained address to contact the MN directly. 

This is the general principle employed by different architectures; it may be 
implemented within different network entities and use very different 
algorithms and protocols. 

\subsection{Presence of Third Distributed Software Entities}

With the aim of managing handovers and the different IP addresses that a MN 
can obtain while moving, several architectural solutions exploit some 
distributed software components in charge of routing data
between the MN and its CN. In the literature, both the term 
\emph{relay} and the term \emph{proxy} are used to refer to these modules.
For the sake of precision, we will use both terms, but not 
as synonyms, in order to emphasize a (maybe subtle but nevertheless important) 
difference.

\subsubsection*{Relay}
It is a software entity, which is mainly used to deal with symmetric NATs or firewalls \cite{STUN,TURN,ice}.

\subsubsection*{Proxy}
It is a software entity that is in charge of routing data/information towards a 
given host, typically in compliance with an application level protocol.

\subsection{Mobility Support and Network Layers}
\begin{figure*}[t]
   \centering
   \includegraphics[width=\linewidth]{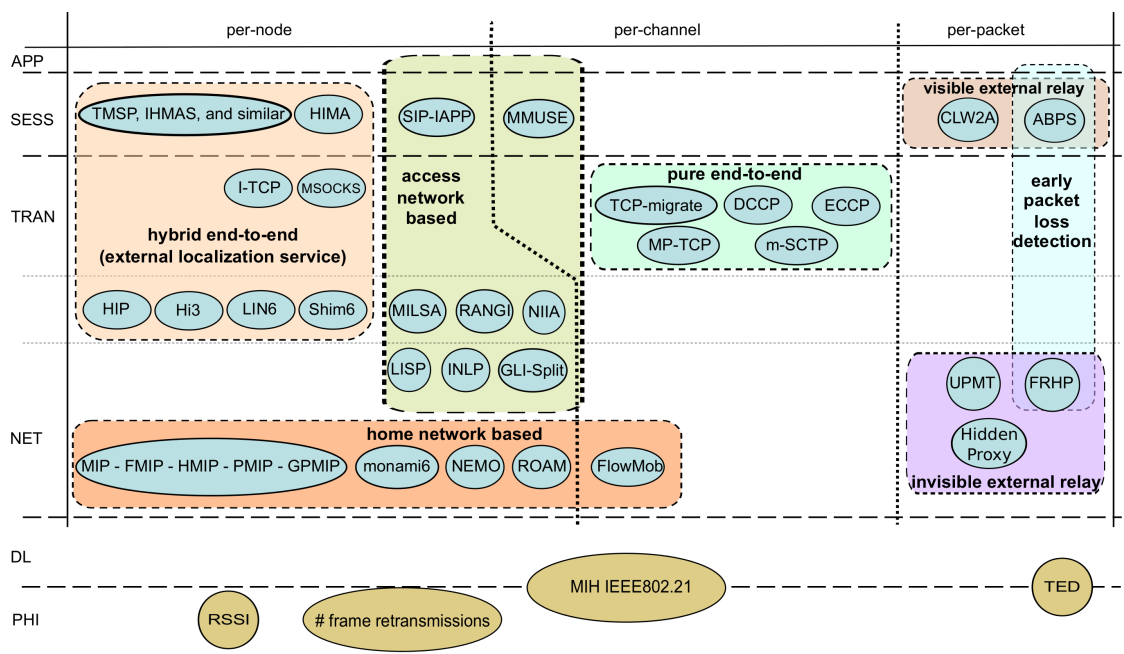}
   \caption{Mobility management architectures}
   \label{fig:architectures}
\end{figure*}

The architectural solutions we will examine in the rest of this survey are 
implemented at different levels of the ISO-OSI reference model, as illustrated 
in Figure \ref{fig:architectures}. This figure illustrates the positioning of the considered systems, 
in the ISO-OSI reference model, starting from the physical layer (see ``PHY'' in the bottom part of the
figure), datalink layer (``DL''), network layer 
(``NET''), solutions between the network and transport layers, solutions 
working at the transport (``TRAN'') and session (``SESS'') layers.
It is worth pointing out that the (brown) circles, placed in the physical and 
datalink layers in Figure \ref{fig:architectures}, are not mobility 
management systems; rather, they are criteria/tools exploited to implement 
mobility management systems. Their description is provided in Section 
\ref{sec:tool_low_levels}.

\subsubsection{Architecture Classification}

The architectures reported in Figure \ref{fig:architectures} can be 
classified depending on where the localization service is deployed (in Figure 
\ref{fig:architectures}, a shaded box groups all systems belonging to the same 
class): 
\begin{itemize}
  \item{\textit{pure end-to-end solutions}} distribute the localization service 
on both the end systems involved in a communication (i.e., MN and CN); 
  \item{\textit{home network-based solutions}} deploy the localization service inside the 
home network to which the MN belongs;
  \item{\textit{access network-based solutions}} (or \textit{border gateway-based 
solutions}) deploy the service inside a border gateway (placed at the edge of 
each network domain), which the MN exploits as its own access point to the Internet;
  \item{\textit{hybrid end-to-end solutions}} manage the reconfiguration on 
both the end systems involved in the communication, but place the 
localization service in a separate server;
  \item{\textit{external relay/proxy solutions}} integrate both the 
localization service and the packet relay service at separate servers, which 
are independent from both the home and access networks. This allows the 
mobility management service to be deployed with no impact on the network infrastructures, and
overcomes the presence of firewalls and NAT systems.  
\end{itemize}

Another important classification relates to the granularity of the service, 
i.e.~to the target (node, channel, packet) to be assigned to a selected NIC. 
(In Figure \ref{fig:architectures}, systems with the same level of granularity 
are grouped in the same column.) When triggered, \textbf{per-node} solutions 
migrate every active flow by adopting a coarse-grained approach and exploiting one NIC 
only, according to the requirements of the whole node. 
Conversely, \textbf{per-channel} solutions allow only one flow to be migrated at a 
time (from one NIC to another) in a finer-grained way compared to the 
previous approach; this enables each flow to exploit its most suitable NIC.
Finally, \textbf{per-packet} solutions route each single IP datagram through 
the most suitable NIC, allowing fine-grained load-balancing and recovery 
policies.  

In the following sections, we shall examine these solutions individually, 
according to the ISO-OSI reference layer to which they belong. We will make a 
distinction between the systems that focus on the use of a single NIC, and 
those that exploit multihoming and the simultaneous use of multiple NICs to enable MNs 
to connect to the network.

\subsection{Tools and Criteria for Mobility Management}\label{sec:tool_low_levels}

As already mentioned, there are some criteria and tools available at the 
physical and datalink layers that can be utilized to implement mobility 
management schemes. This subsection will give a brief outline of each of them.

The Received Signal Strength Indicator (RSSI) is a measurement of 
the power level being received by the antenna of the NIC. Thus, the higher the 
RSSI the stronger the signal. Signal strength is one of the main metrics a MN can 
use in order to discriminate amongst different APs and select the best one to 
connect to.

Another important criterion is the number of frame retransmissions needed at the lower (Data-Link) layers 
to deliver a frame from the MN to its AP. Mobility management systems can use this metric as a handover decision criterion.

802.21 is an IEEE standard for dealing with seamless handovers in heterogeneous 
networks. It has been devised for supporting vertical handovers or, as 
called in the standard, Media Independent Handovers (MIH).
It defines a set of handover-enabling functions to assist MNs during the 
handovers and handover decision making. Thus, these primitives offered at the operating system level can be exploited to build mobility management architectures.
ODTONE is an IEEE 802.21 implementation that is operating-system-independent and open source \cite{802.21imp}.

Finally, the Transmission Error Detector (TED) is a software tool that provides 
the MN with information about successful (unsuccessful) datagram reception 
at the AP \cite{GhiniJSS}. 
This information can be delivered to software modules working at higher 
(i.e.~application) layers of the protocol stack. In this way, it is possible to 
devise cross-layer strategies for managing NICs and performing vertical 
handovers.

\subsection{Coping with NAT and Firewall Systems}

Before going into the details of the systems available in the literature, we shall discuss the possible presence of firewall and NAT systems placed in between end-nodes involved in a communication. 
This is a fundamental issue and one that is often ignored by many mobility management solutions.

Network Address Translators (NATs) are common devices that ``hide'' private 
networks behind public IP addresses.
A NAT device works by associating a public address and port with a private 
destination address and port. In essence, the host has a private (and local) 
transport identifier, which is different from the one that is seen outside the private 
network. The NAT is responsible for translating the private 
address into the public address (for outgoing messages) and vice versa (for 
incoming messages). As a consequence, hosts within a private network behind a 
NAT are allowed to initiate a communication with a host outside the private 
network, but quite often the reverse is not possible. Indeed, the CN may not 
know the public transport identifier (that can be created in real-time by 
the NAT and change in time).
Similarly, many firewalls allow connections to be initiated from the private 
network only, with the effect that these nodes cannot be contacted from other nodes outside the private network
to initiate a new communication. 

This common practice represents an important limitation for the support of 
mobility.
To cope with this possible situation, the applications can resort to services offered 
by external \ac{STUN} \cite{STUN}, \ac{TURN} \cite{TURN} and \ac{ICE} \cite{ice} systems. These 
are protocols that allow a host (a MN in our case) to learn its public address, 
as it is seen outside the private network in which it is located.
In this way, the MN can activate procedures that enable external hosts to contact 
it directly, e.g.~it can expose its public address in some discovery service.  

Unfortunately, the TURN, STUN, ICE protocols do not work correctly when both the 
end systems (MN and CN) are protected by (the most common and restrictive) 
symmetric NATs and firewalls. 
In this case, each connection of a MN with other nodes is treated separately; 
moreover, a node does not have a bijective mapping between a local 
transport-layer address and a public address.
This represents a problem, since a typical approach in network programming is to 
have nodes that 
employ the same $\langle\text{IP address, port}\rangle$ transport address for 
establishing different transport-level communications (e.g., a node exposes its 
transport address $\langle\text{IP address, port}\rangle$ waiting for new 
connections).

However, the symmetric NAT/firewall maps every communication that has the same internal IP address and port number, locally at the MN, to a different external transport id. 
Thus, it is not possible to expose a $\langle\text{IP address, port}\rangle$ 
pair to all possible CNs, since
only the CN that receives a packet from a given external transport id representing 
the MN can send a packet back. 
The problem is that if each end-system is behind a symmetric NAT, then 
each system cannot obtain an address to connect to its remote correspondent, as the necessary external transport id of this correspondent has not yet been created by its relative NAT/firewall.

\begin{figure}[h]
   \centering
   \includegraphics[width=.6\linewidth]{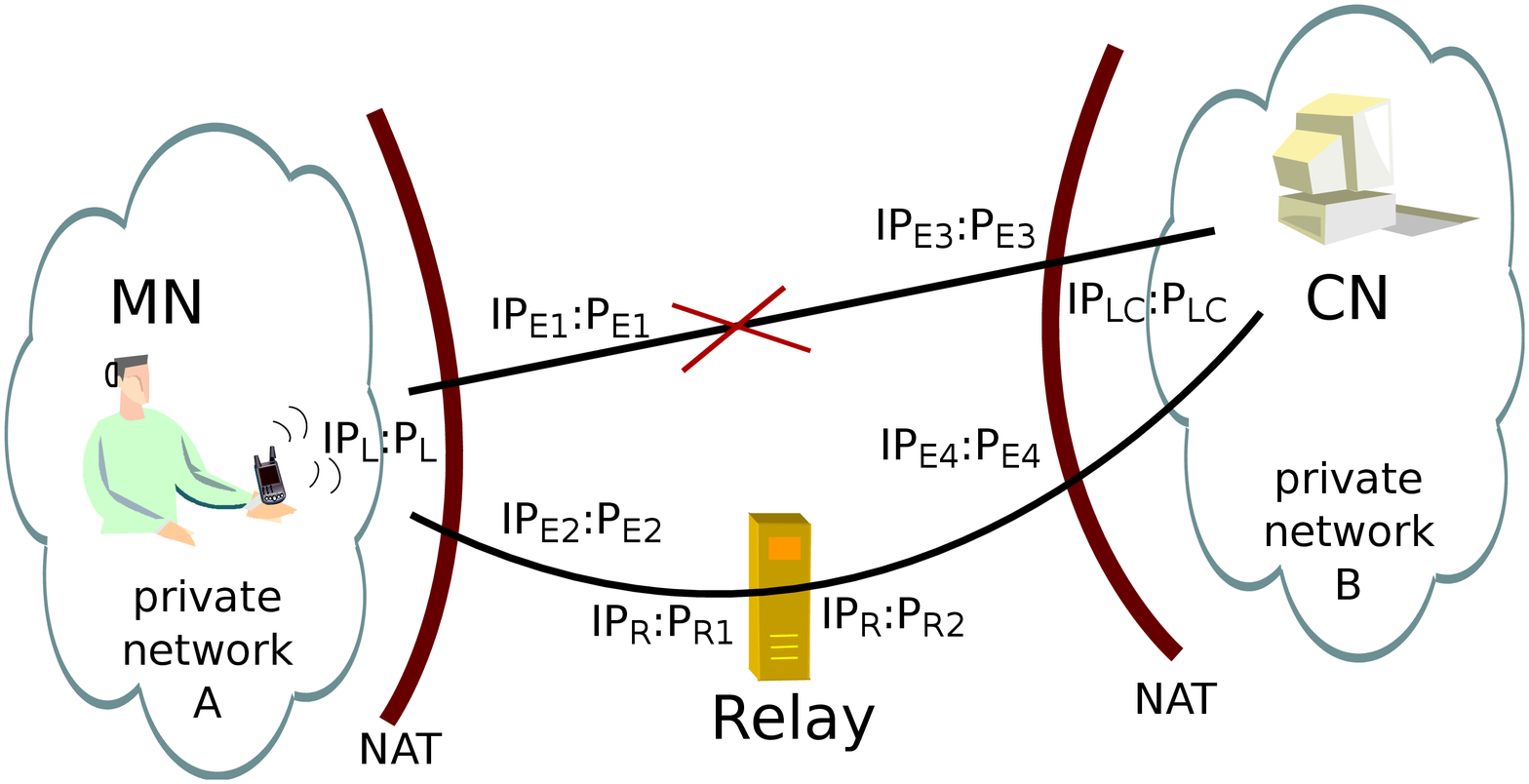}
   \caption{The data relay allows the undirected communication between two nodes 
behind different firewalls or NAT systems}
   \label{fig:nat}
\end{figure}

Basically, these restrictive firewalls require an intermediate 
application-layer relay server, outside any firewall and NAT systems, that 
receives and forwards the packet exchanged between the MN and its CN for the 
entire duration of the communication (see Figure \ref{fig:nat}). The relay 
rewrites the transport and network datagram header so as to appear as the peer 
node for both MN and CN.
In some cases, it is even necessary to rewrite some data inserted 
into the payload at the application-level. 
For instance, SIP inserts information on the end point, related to the network and transport levels, within the application payload (e.g.,~the INVITE message 
contains the IP address and port of the end-point). Thus, when an end-point is 
behind a NAT, this information must be modified (we discuss this issue in Section \ref{sec:session}). 
The problem is that, as already mentioned, if the MN is behind a symmetric NAT/firewall, 
the external IP address and port are assigned only when the application data transmission begins; 
thus, the INVITE message cannot contain such information
unless an external transport id is inserted (i.e.~the one offered by the relay).

\begin{figure}[H]
   \centering
  \subfigure[no firewalls] {
    \includegraphics[width=.45\textwidth]{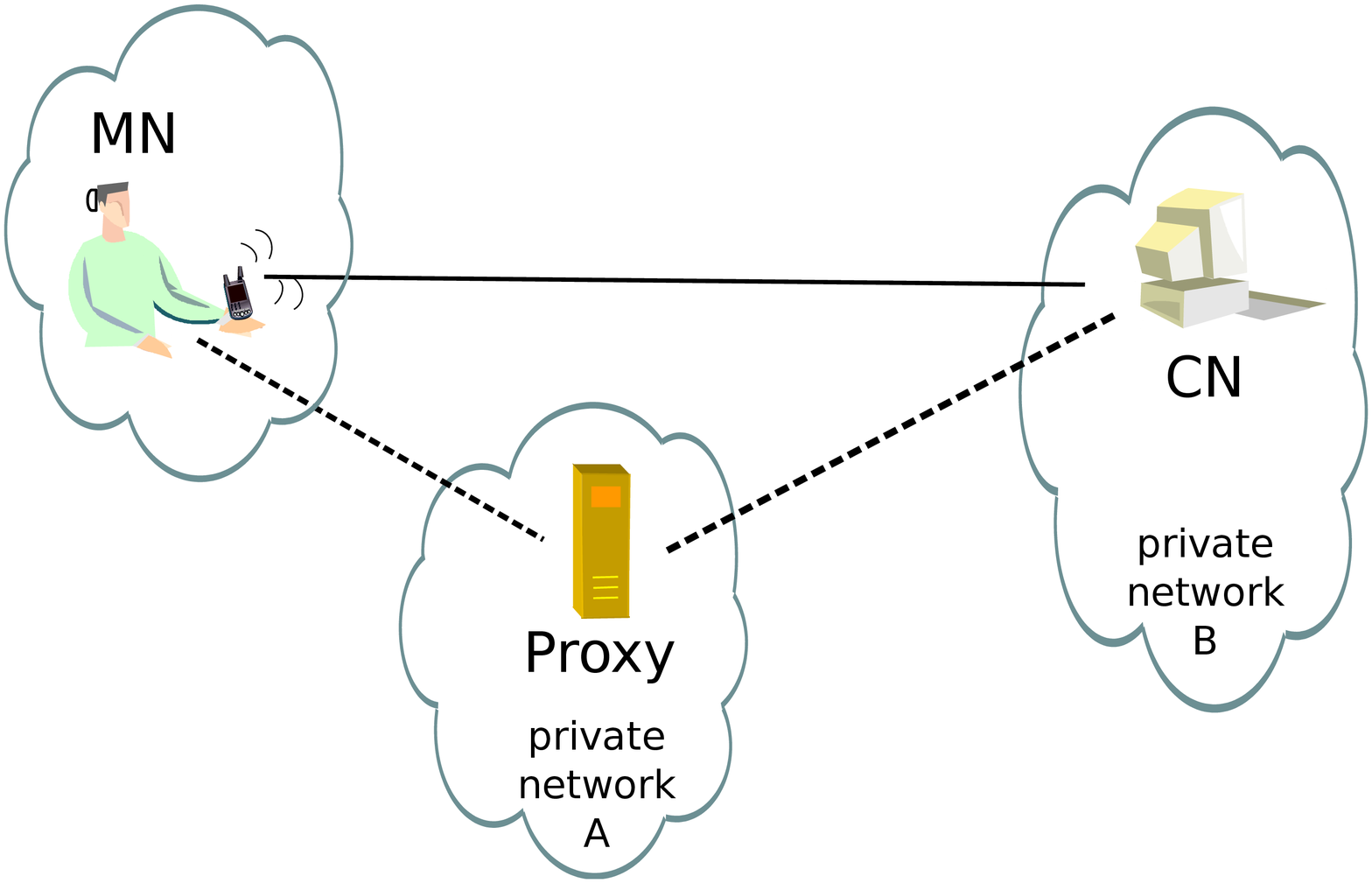}
    \label{fig:mipv6_fw:no_fw}
  }
  \subfigure[Proxy behind a firewall (impractical situation)] {
    \includegraphics[width=.45\textwidth]{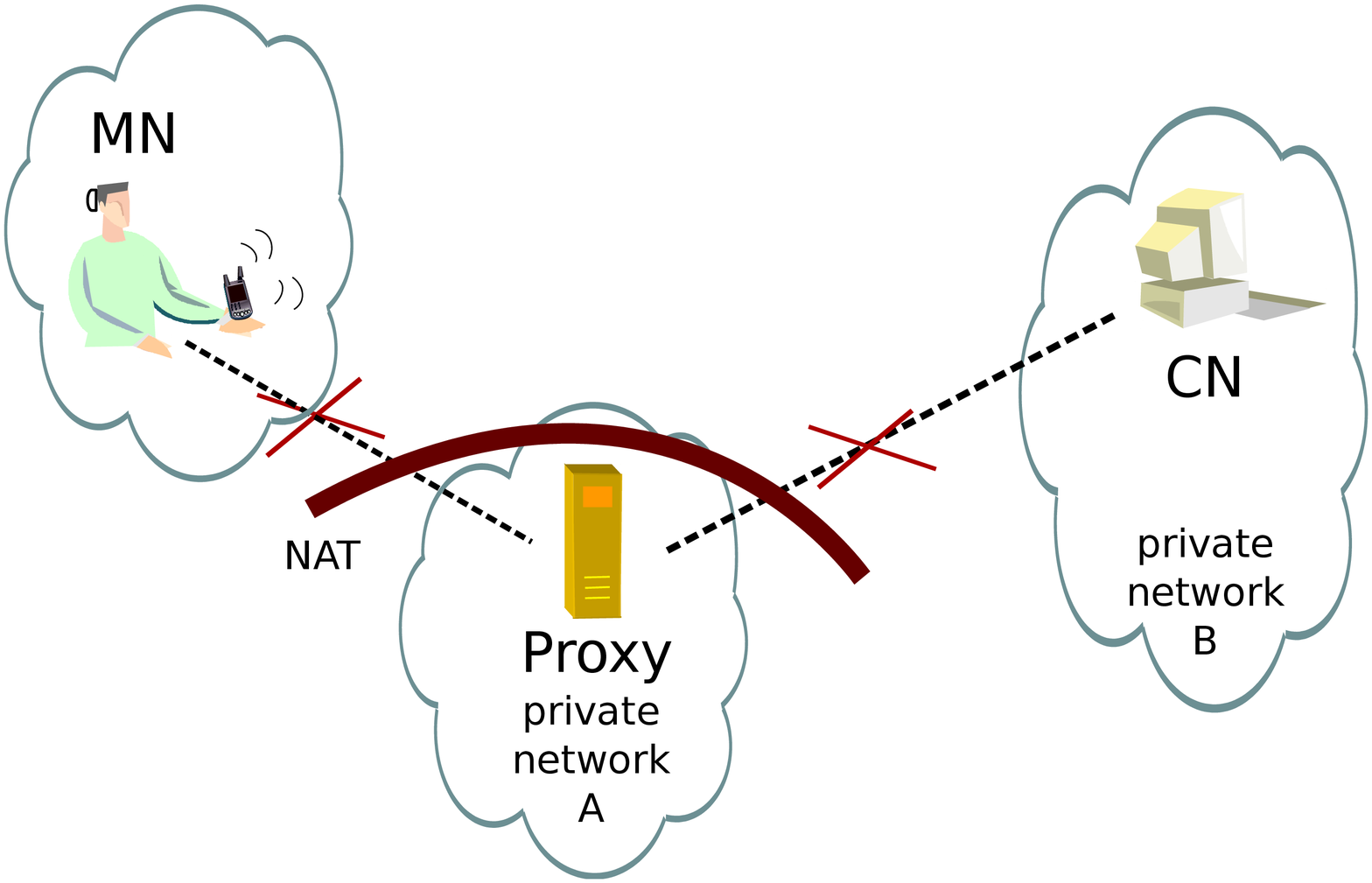}
    \label{fig:mipv6_fw:ha}
  }
  \subfigure[Proxy behind a firewall - use of a relay (impractical situation)] {
    \includegraphics[width=.45\textwidth]{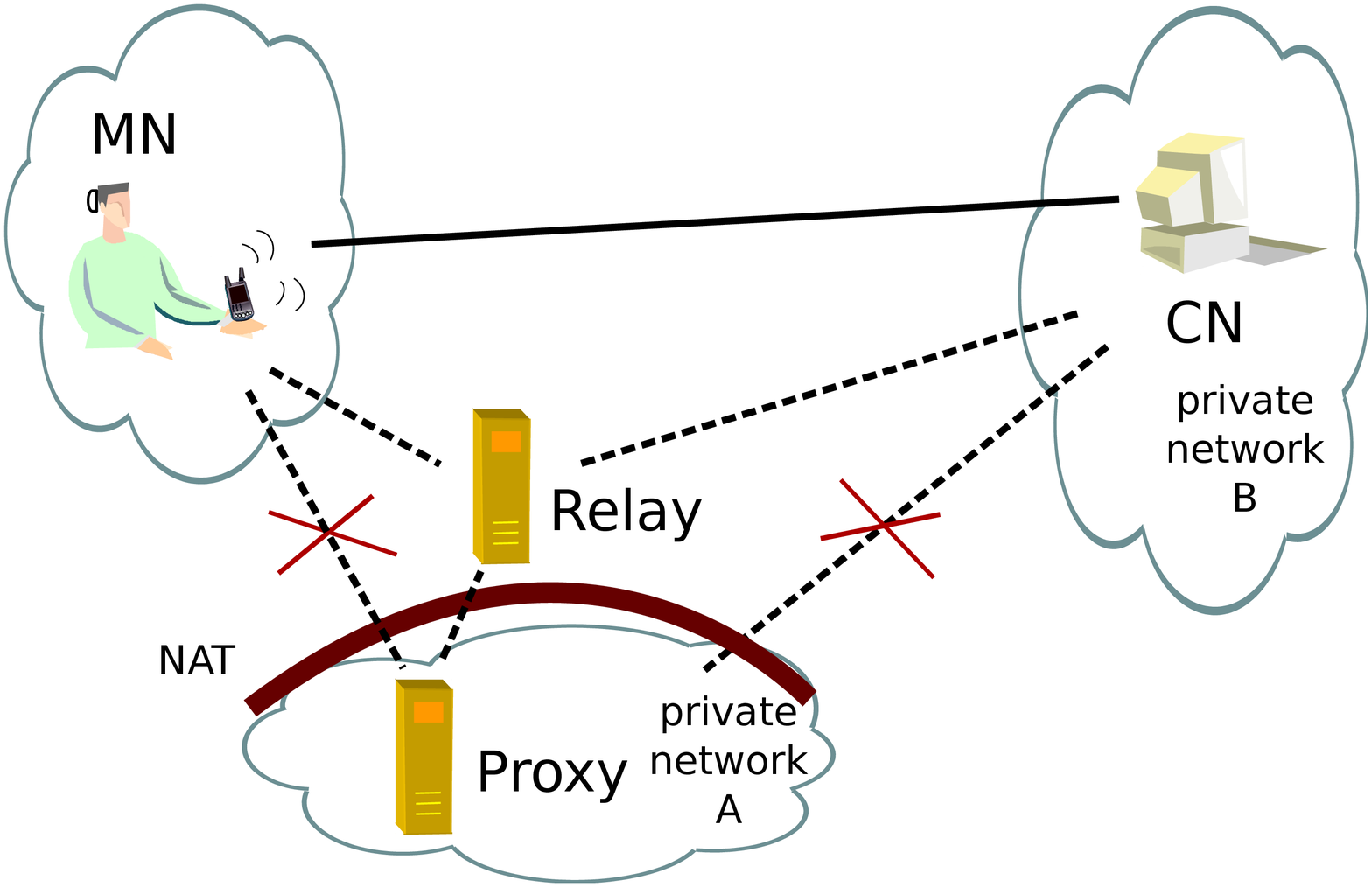}
    \label{fig:mipv6_fw:ha_rel}
  }
  \subfigure[Proxy with relay] {
    \includegraphics[width=.45\textwidth]{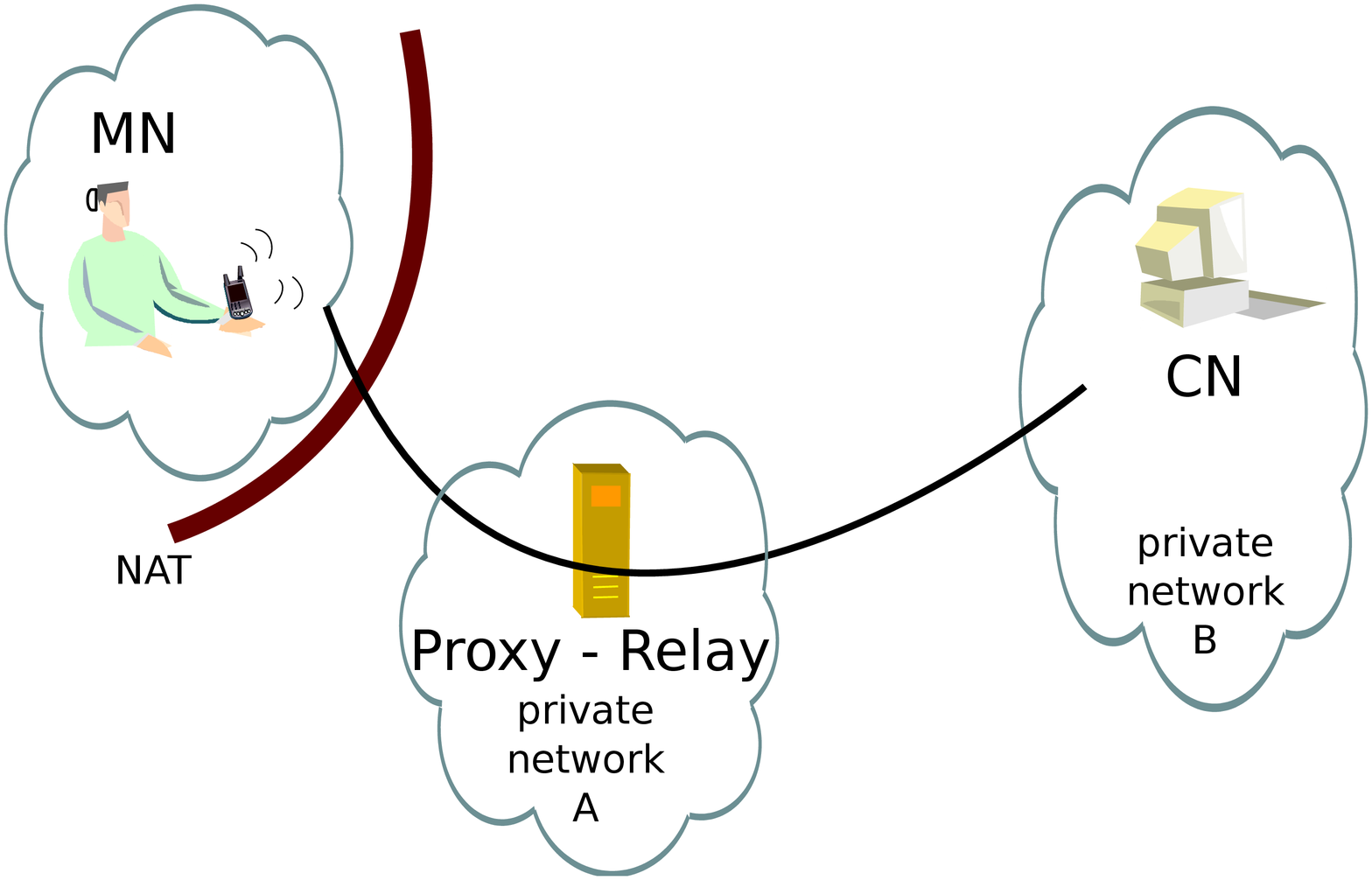}
    \label{fig:mipv6_fw:mn}
  }
  \subfigure[CN behind a firewall] {
    \includegraphics[width=.45\textwidth]{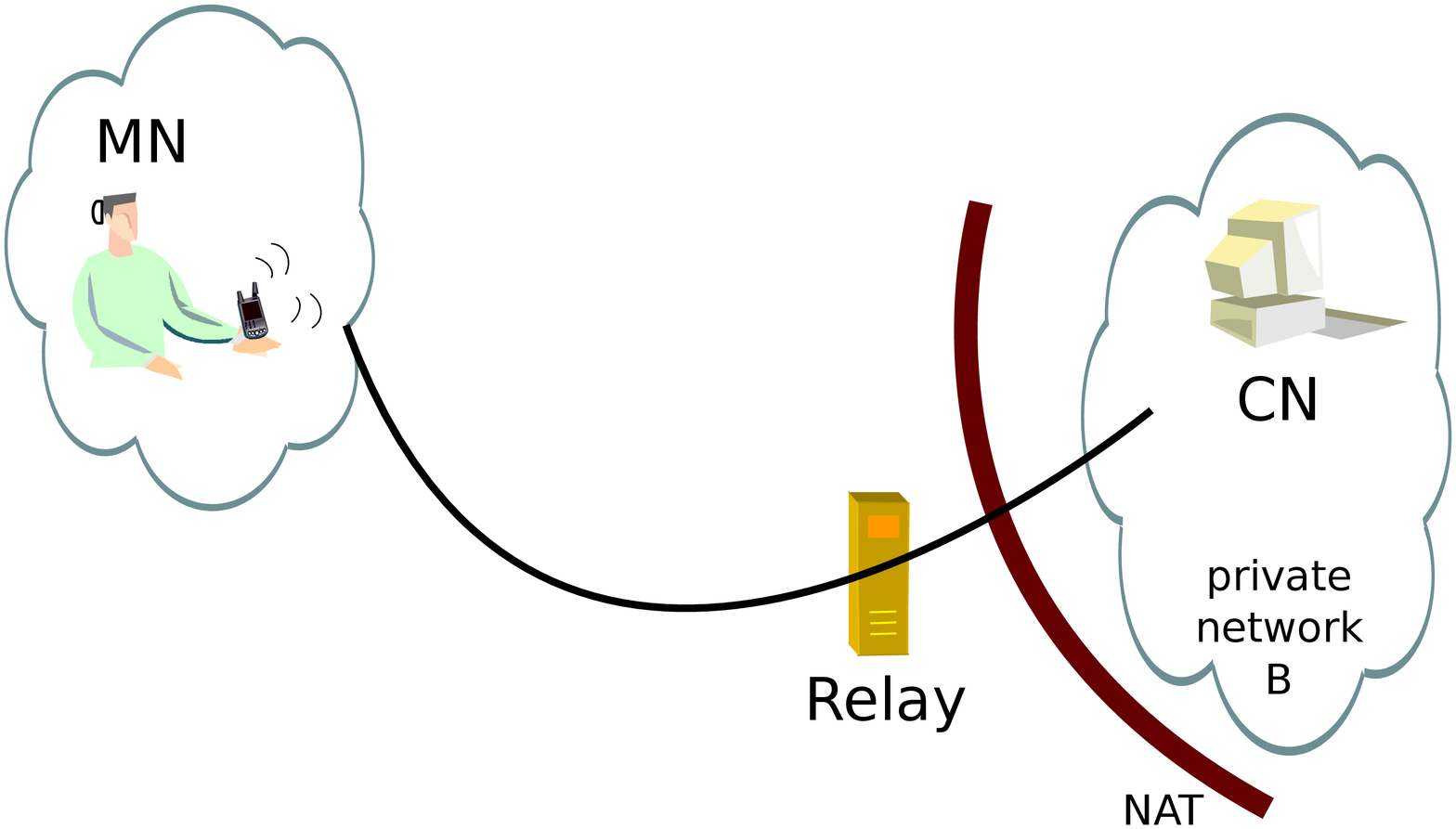}
    \label{fig:mipv6_fw:cn}
  }
  \subfigure[MN and CN behind a firewall] {
    \includegraphics[width=.45\textwidth]{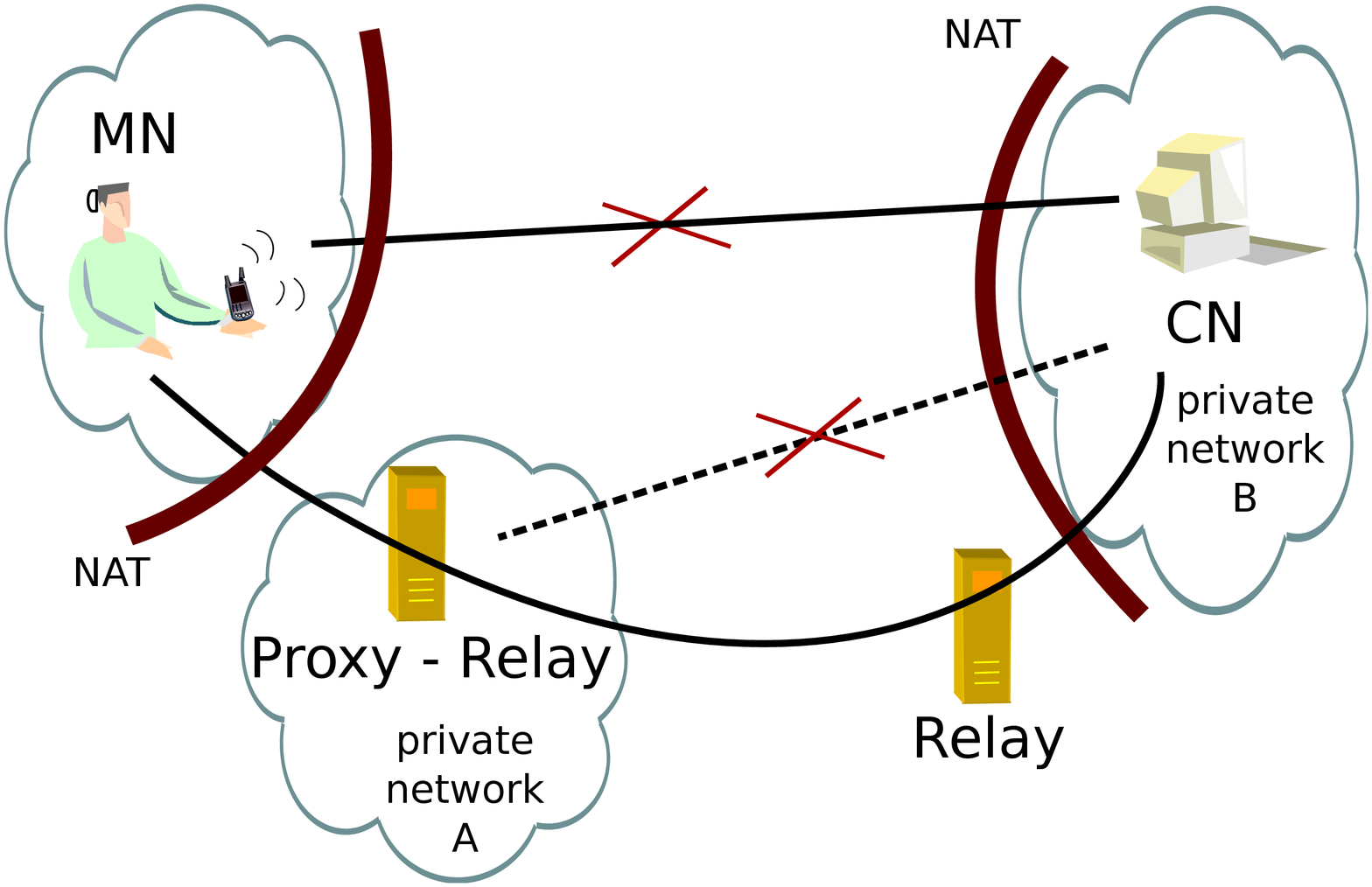}
    \label{fig:mipv6_fw:mn-cn}
  }
  \caption{Presence of symmetric NATs/firewalls and need for a 
relay/proxy. The proxy refers to hosts that enable MNs to 
communicate while moving (e.g.~the Home Agent in MIPv6).}
  \label{fig:mipv6_fw}
\end{figure}

Figure \ref{fig:mipv6_fw} shows the possible cases that may arise during a 
communication between a MN and its CN. 
The main conclusion is that a relay node is needed if we want to guarantee mobility support in any given case. In Figure \ref{fig:mipv6_fw}, the proxy represents the software entity which is in charge of managing the MN mobility, provided that the system under consideration uses such a software component. 
The name ``proxy'' is generic and it is used in several works, e.g.~SIP-based 
approaches \cite{RFC3261,Bellavista:2010,Kalmanek:2006,Schulzrinne:2000,
Udugama:2007}, ABPS \cite{GhiniJSS}. As to other approaches, such as MIPv6-based approaches (see Section \ref{sec:mipv6}), the proxy in 
the figure represents the \acf{HA}.

As shown in Figure \ref{fig:mipv6_fw}, a MN and its CN can communicate directly 
when there is no NAT/firewall constraining the communication (Figure 
\ref{fig:mipv6_fw:no_fw}).
Then, the figure reports the impractical case when the proxy is behind a 
symmetric NAT/firewall (Figure \ref{fig:mipv6_fw:ha}). In this case the MN and 
the CN cannot contact the proxy; as a result, no communications can be established.  Inserting a relay for the proxy might be a solution (Figure 
\ref{fig:mipv6_fw:ha_rel}), but this would certainly be an unusual situation: if we 
install a proxy in a network, we should be aware that this entity must be 
reached by hosts outside the private network. Hence, the system designer must 
take this factor into account.

A more realistic scenario is when the MN connects to the Internet via a private 
network, which is behind a NAT/firewall (Figure \ref{fig:mipv6_fw:mn}). In this 
case, the MN node can contact the proxy that would act as a relay if the CN asks 
to connect to the MN.
The opposite occurs when the CN is behind a NAT/firewall 
(Figure \ref{fig:mipv6_fw:cn}). 
In this case, a relay is required to let the MN establish a communication. Indeed, the MN cannot connect to the CN directly, because of the NAT/firewall. The same occurs if a proxy is running for the MN. Moreover, the CN has to declare that a connection is possible through a relay, at a certain transport address.
This is necessary, otherwise the MN is not able to initiate a communication with the CN. 
However, it is important to point out that in the two last cases, the 
application must be aware that a relay node has to be contacted; thus,
support at the application level is needed to enable the two nodes to interact. 

Finally, a further case arises when both MN and CN are behind a NAT/fire\-wall 
(Figure \ref{fig:mipv6_fw:mn-cn}); then, both end-systems must employ a relay to allow their correspondent to contact it.

\subsection{Implications on the Use of Session and Application Layer Protocols}
\label{sec:session}

The considerations reported in the previous subsection suggest that, under 
certain circumstances, if we want to add mobility and multihoming to current 
Internet applications,
some implementation tricks are needed at the session and application layer 
protocols. (Hereinafter, we will refer to all protocols working over the 
transport layer as ``application protocols''). There are in fact
application-level protocols that
\begin{enumerate}
 \item do not respect the protocol stack stratification. They typically insert some network and transport layer data into the application payload. There are numerous applications included in this group. For instance, all applications exploiting SIP have these characteristics;
 \item respect the separation of roles of protocols in the network stack. Thus, 
each application message implicitly exploits ports and IP addresses specified 
in the lower level protocols. HTTP-based applications are examples of this 
type. 
\end{enumerate}

The presence of lower level data in the application payload (first case) represents a problem, firstly, when end-points are behind a NAT or firewall, and secondly, for building a mobility architecture that supports multihoming. Indeed, in the latter situation,
there are multiple IP addresses (referring to 
different NICs) to be managed at the network and at the application layers. The 
CN might receive different application messages from the MN through multiple IP 
addresses. 
Mobility architectures need to cope with this issue. 

There are basically two kinds of solutions, which will be presented in the next 
sections.
One refers to the idea of separating the notion of ``node identifier'' from the 
``locator'' of that node. This type of solution is devised for the so called 
``Future Internet''. Another type of solution requires a 
proxy. 
Examples of this second kind are ABPS \cite{GhiniJSS} and all MIPv6-based 
approaches \cite{mipv6}, where the \ac{HA} (which is a proxy) is responsible for 
routing messages between the MN and its CN; hence, the CN sees the address of the 
\ac{HA} only as its interlocutor.
When applications exploit proxies already (e.g.~HTTP and SIP-based 
applications), this solution has a simpler implementation, since the proxy can 
be upgraded to incorporate the functionalities needed to support multihoming.

\section{Single NIC-based Architectures}
\label{sec:singleNIC}

This section focuses on architectural solutions based on the use of a 
single NIC to manage mobile communications. Table \ref{tab:s_nic} summarizes
all the techniques described below, classified according to the protocol layer they 
operate on.

\begin{table*}[t]
\caption{Single NIC-based Architecture -- Protocol Stack Level Classification}
\label{tab:s_nic}
 \centering
 \scriptsize
 \begin{tabular}{|l|C{10cm}|}
 \hline
 \hline
  {\bf Level} & {\bf Approaches} \\
 \hline
 \hline
 Session & SIP-IAPP \cite{WuYH07} \\
 \hline
 Network & MIPv6 \cite{mipv6}, FMIP \cite{VanHanh:2008}, HMIP \cite{rfc-4140}, 
           PMIP \cite{rfc5213}, GPMIP \cite{Zhou:2010}, \cite{He:2010}, 
           NEMO \cite{Perera:2004}, LISP \cite{lisp}, ROAM \cite{Zhuang:2005}\\
 \hline
 \hline
\end{tabular}
\end{table*}

\subsection{Solutions at the Network Layer}

\subsubsection{Mobile IPv6}\label{sec:mipv6}
Amongst the architectural solutions working at the network layer, it is 
worth citing the efforts in the Mobile IP version 6 (MIPv6) 
\cite{mipv6} and its optimizations, e.g. the Fast Handover Mobile IPv6 (FMIP) 
\cite{VanHanh:2008}, Hierarchical Mobile IPv6 (HMIP) \cite{rfc-4140},  Proxy 
Mobile IPv6 (PMIP) \cite{rfc5213}, GPMIP \cite{Zhou:2010}, and \cite{He:2010}. 
All these approaches employ a \acf{HA}, i.e. a proxy working inside the access 
network to which the MN belongs. The \ac{HA} plays the role of the location 
registry, and routes datagrams towards the MN when this node is outside its 
``home network''. 
In addition, if the MN wishes to register its binding with a CN, so as to 
communicate directly with it, without the interposition of the \ac{HA}, that MN 
must perform return routability operations \cite{mipv6}. 
In order to work properly, MIPv6-based approaches need all the end-systems to have IPv6 capabilities so as to insert some extension headers that transport both the MN’s identifier (the home address) and the current MN address into the IP datagrams.

A clear limitation is that these architectural solutions only work on infrastructures 
with IPv6 capabilities. 
Moreover, the MIPv6 specification does not allow the simultaneous use of the 
multiple MN NICs. 
For each given MN, the address of a single NIC is registered at the \ac{HA}. In 
addition, as demonstrated in \cite{kong}, the handover latency is very high 
due to the numerous authentication messages, which causes a service disruption 
time that is not compatible with strongly interactive services such as VoIP. 

\subsubsection{NEMO}
Network mobility Basic Support Protocol (NEMO BSP or simply NEMO) 
is concerned with managing the mobility of an entire network. NEMO 
aims at providing seamless Internet connectivity of the whole mobile network 
that consists of Mobile Routers (MRs) and \acp{MN}. 
The application scenario is public transportation, such as trains and buses 
\cite{Perera:2004}. The network moves around as a whole, along with vehicles. 
NEMO BSP is based on MIPv6 with minimal extensions. Therefore, the handover 
mechanism of a MR is essentially the same as that of MN in MIPv6. 
In NEMO BSP, a MR serves as a gateway; a permanent address called Home Address 
(HoA) is obtained on the home link as an identifier of the MR. When the MR moves away, 
it acquires a \ac{CoA} from the access router in the foreign network. MR 
sends a ``binding update'' message to its \ac{HA} proxy located in the home 
network, binding the \ac{CoA} with the HoA. 
After the binding process, a bi-directional tunnel is established between the MR 
and the HA proxy. Packets from the CN with the destination of MR’s HoA are 
directly routed to the HA, and the HA is in charge of rerouting all packets to 
the \ac{CoA} of the MR through a tunnel. 
MNs in the mobile network have permanent addresses taken from the mobile 
network prefix advertised on the MR’s ingress interface, and packets intended 
to or originated from the MNs are encapsulated in the tunnel.

\begin{figure}[h]
   \centering
   \includegraphics[width=\linewidth]{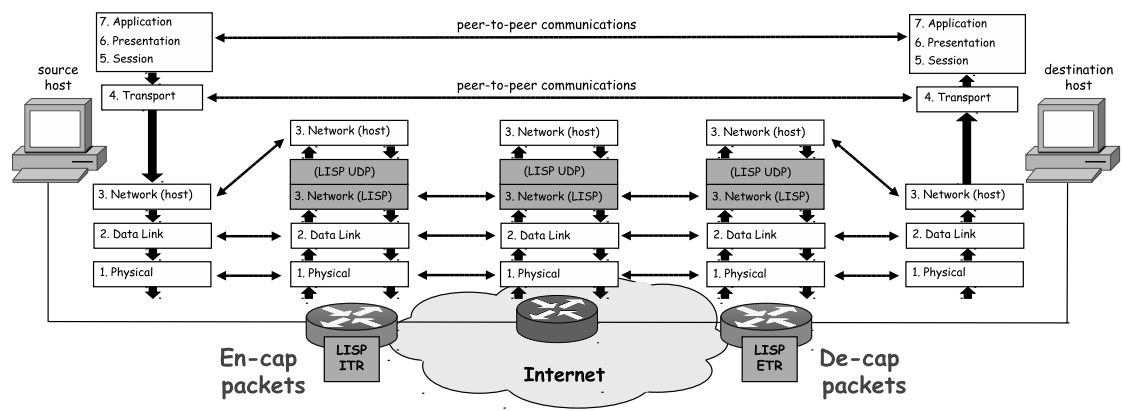}
   \caption{The LISP architecture \cite{lisp}}
   \label{fig:lisp}
\end{figure}

\subsubsection{LISP}
Another noteworthy architecture that belongs to the border-gateway class is the Location/ID Separation Protocol (LISP) 
\cite{lisp} proposed by Cisco. This is 
a first solution that produces the aforementioned separation between the identifier 
of a node, and its current location in the network.
As depicted in Figure \ref{fig:lisp}, LISP makes use of an overlay network of 
LISP routers, located at the edge of the network domains and classified as 
Ingress Tunnel Router (ITR) and Egress Tunnel Router (ETR).
The ITR intercepts the IP datagrams coming from the MNs, maps the sender address 
(the location) to the sender ID, encapsulates these datagrams in LISP packets 
and routes them through the LISP routers towards the destination. 
When a LISP packet reaches the ETR at the edge of the destination domain, the 
ETR extracts the IP datagram from the LISP packet, maps the destination ID to 
the destination locator and routes the IP datagram to the destination. 
In that way, LISP does not require changes to the end systems. 

The main drawback of LISP is shared with all other architectures that introduce functionalities at the edge of the network domains, i.e. they need all the domains to be LISP enabled by deploying Ingress/Egress tunnel routers at their edges. 
Indeed, this architecture needs all the paths between each MN and the Internet to flow through at least a LISP-enabled border router. If a MN’s NIC connects to a network domain that has no LISP-enabled router at its edge, the LISP architecture fails to provide network continuity to the MN.
Recent proposals have been published that provide functionalities for NAT 
traversal for LISP MNs \cite{Klein:2010,ttr}.

\subsubsection{ROAM}
Some proposals at this level exploit the service provided by Internet Indirection
Infrastructure (i3)
\cite{Stoica:2004}. i3 builds a rendezvous-based communication abstraction. The 
system extends point-to-point communications by decoupling the act of sending 
and the act of receiving messages. The main aim is to provide an approach to 
ease the development of multicast communications.

In i3, a logical identifier is associated to each node. This 
identifier can be mapped to the current IP address by means of an indirection 
point. In this way, when we want to send a message to a CN, the message passes through an 
i3 server to locate the destination current address.

ROAM \cite{Zhuang:2005} exploits and extends i3 to support node mobility. In 
substance, the indirection point is exploited to manage handovers. 
A proxy-based solution is adopted to transparently support unmodified 
applications on mobile nodes. Moreover, the use of an i3 server guarantees that 
communication is established also in the event of simultaneous handovers of the two mobile nodes. 

Support for legacy applications is guaranteed by the introduction of a proxy. However, any application host is required to have a related proxy. Hence, each server should deploy an i3 proxy in order to communicate with the MN. This represents an obvious limitation for the deployability of the system in the current Internet.

Finally, it is worth pointing out that this solution needs a non-negligible amount 
of additional information to be inserted into packets. Thus, header compression is 
needed to reduce the packet header overhead.

\subsection{Solutions at the Session Layer}

When solutions are devised to let nodes communicate through different networks, 
a key role might be played by session protocols that control the dialogue 
between end-points and incorporate functionalities used by the localization 
service. 
Today, Session Initiation Protocol (SIP) is the main protocol employed for 
controlling multimedia, multi-homed communication sessions 
\cite{RFC3261,Schulzrinne:2000}. We now review its main characteristics.

\subsubsection{Session Initiation Protocol (SIP)}

SIP is a session-layer text protocol that uses a message/response handshake for 
signaling purposes. In particular, it is used to establish or change 
communication parameters such as IP addresses, protocol ports and audio/video 
codecs between the end-systems. The SIP specification is extensible and allows 
application-defined fields to be added to the SIP messages. 

The SIP messages worthy of mention here are REGISTER, INVITE and re-INVITE. 
The REGISTER message allows a given node to declare that it is available for 
communications; it is usually sent to a SIP server that works as the 
localization service. The INVITE message is used to establish a communication 
session between two nodes. Typically, this message is sent from the user to the 
SIP server that replies with the address of the other end node, together with 
some communication parameters. 
Then, the two end nodes can communicate directly. A re-INVITE message may be 
used when communication parameters (such as the IP address) change. 

Another important aspect is that the SIP protocol allows the presence of SIP 
proxies, that can be transparent to the application (proxy agent) or can 
masquerade the end-systems (Back-to-Back (B2B) user agent) working as an opaque 
relay.

\subsubsection{SIP and IAPP-based proposals}
A SIP-based mechanism that focuses on WiFi technologies has been proposed that 
extends the already mentioned IAPP \cite{WuYH07}. 
This approach (hereinafter called SIP-IAPP) proposes a cross-layer approach that modifies the IAPP to speed up the handover in WLANs. This allows direct forwarding of messages addressed to the MN, from an old AP to the new one, while the MN is still reconfiguring its SIP network setting.

This enhancement speeds up the SIP registration, increasing the speed of the handover process as well. However, these approaches again require modifications in the infrastructure, and focus on a single networking technology.

\section{Architectures that Enable Multihoming}
\label{sec:multihoming}

This section reports solutions that cope with handovers by resorting to the 
possible use of multiple NICs. These schemes work at different levels of the 
network stack.
Table \ref{tab:m_nic} summarizes all the techniques described in this section, classified based 
on the protocol layer they operate on.

\begin{table*}[t]
\caption{Multihoming Architecture -- Protocol Stack Level Classification}
\label{tab:m_nic}
 \centering
 \scriptsize
 \begin{tabular}{|l|C{8cm}|}
 \hline
 \hline
  {\bf Level} & {\bf Approaches} \\
 \hline
 \hline
 Cross-Layer & ABPS \cite{GhiniJSS}, CLW2A \cite{FerrettiG09} \\
 \hline
 Session & IHMAS \cite{Bellavista:2010}, TMSP 
	   \cite{LimYLL09}, \cite{Udugama:2007}, \cite{Kalmanek:2006},
	   MMUSE \cite{SalsanoPMNV08}\\
 \hline
 Transport & DCCP \cite{rfc4340}, m-SCTP \cite{Budzisz:2012}, TCP-migrate 
	     \cite{snoeren2001reconsidering}, MPTCP \cite{mptcp,Paasch:2012}, 
	     MSOCKS \cite{msocks}, I-TCP \cite{itcp}, ECCP \cite{eccp}
 \\
 \hline
 Between Network and Transport & HIP \cite{BokorZNJ09,rfc4423}, 
	      Hi3 \cite{GurtovKLN08}, LIN6 \cite{draft-teraoka-ipng}, 
	      MILSA \cite{Pan:2010}, NIIA \cite{niia,Schutz:2010}, 
	      RANGI \cite{rangi}, 
	      Shim6 \cite{rfc5533} \\ 
 \hline
 Network & monami6 \cite{li07,rfc4861}, FlowMob \cite{Toseef:2008}, 
	   GSE \cite{draft-ipng-gseaddr-00.txt},
	   ILNP \cite{rfc6740}, GLI-Split \cite{gli-split}, 
	   hidden proxy \cite{ghi06}, UPMT \cite{Bonola:2009},
	   FRHP \cite{Giordano:2012} \\
 \hline
 \hline
\end{tabular}
\end{table*}

\subsection{Solutions at the Network Layer}

\subsubsection{Monami6}
An extension of MIPv6, called Multiple Care of Address registration (monami6) 
\cite{li07,rfc4861,Wakikawa03multiplecare-of-address,pan2008,sousa11} has been 
proposed for supporting host mobility and multihoming. 
If a MN configures several IPv6 global addresses on one or more of its 
NICs, it can register these addresses with its \ac{HA} as \acp{CoA}. This 
enables the \ac{HA} to forward messages to the MN.  

As for many other solutions, this extension does not overcome
firewall and NAT systems, since according to IPv6 based approaches, with the 
advent of IPv6 there will be no need to employ NAT technologies anymore. 
In addition, the return routability operations cannot be easily extended to 
verify multiple  \acp{CoA} and, as usual, if the CN is protected by a firewall 
(or NAT system), the return routability operation fails \cite{Chen2010407}.

\subsubsection{FlowMob}
The Flow Mobility technique (FlowMob), described in \cite{Toseef:2008}, uses a 
multiple \ac{CoA} to enable the MN to register its multiple IP addresses with its 
\ac{HA}. 
This extension allows the MN to separate its outgoing traffic into different flows 
based on the protocols, port numbers and IP addresses, and forward each given 
flow using a selected NIC. 
This technique allows a flow-based switching through the MN’s NICs. In 
\cite{ZafeirisG11}, the problem of assigning traffic flows to available 
interfaces is optimized using a heuristic algorithm. 

This approach shares the same limitations as the other MIPv6-based approaches: 
it needs the network infrastructures to be modified in order to add IPv6 
capabilities. Moreover, the handover latency is very high, due to the 
large number of authentication messages. 
Finally, as mentioned for other IPv6 based strategies, this approach does not 
provide solutions for symmetric firewalls, and thus requires an external relay 
to be used.

\subsubsection{Separation between locator and identifier; ILNP}
A completely different scheme refers to the already mentioned idea of modifying 
the Internet architecture to perform the mapping between the identifier and the 
locator of a node. One of the first proposals in this direction was the Global, 
Site, and End-system address elements (GSE) \cite{draft-ipng-gseaddr-00.txt}.
A prominent example of approaches devised to support mobile nodes is 
the Identifier Locator Network Protocol (ILNP), where the DNS and ICMP protocols 
are modified/updated in order to support the possible changes of the locator 
for a given identifier \cite{rfc6740}. 
The approach requires a novel implementation of the network and transport layers 
at the nodes, while it is compatible with the IPv6 backbone. 

Basically, when a MN is moving between two distinct networks, it updates 
its locator record in the DNS. Thus, if new sessions are established, they are 
established directly to the node's current location.
As regards active sessions, the approach enables the MN to inform its CN directly of 
the changes, using newly defined ICMP Locator update messages. 

This approach enables multihoming. Moreover, the 
presence of NATs does not represent a problem, provided that applications employ 
identifiers only, and no locators, which means that they employ some \acp{API} that do not exploit 
lower level information (to be implemented in accordance with the new protocols) \cite{Atkinson:2007}.

Although simple and elegant, these solutions have been devised to work in future Internet 
scenarios. Thus, this type of approach requires modifications of the 
infrastructure and its protocols. 
Moreover, updates of locators at the DNS are not instantaneous; in addition, given the amount of mobile nodes that are on the Internet nowadays, this solution presents some scalability problems (that can be easily solved, but which do have some cost). Hence, this approach may cause some service 
unavailability, when message updates are lost and both end-points are 
mobile.

\subsubsection{GLI-Split}
The ``Global Locator - Local Locator - Identifier Split'' (GLI-Split) is a system that
adds a global mapping service to the IPv6 architecture. Locators are 
distinguished between local ones, employed for local routing within an edge 
network, and global ones \cite{gli-split}. The mapping system tracks changes 
for the locators, thus enabling mobility and multihoming. 
The architecture employs border routers at the edge networks that are in charge 
of modifying the address information (from local to global and vice versa)
contained in packets traversing different edge networks.
The system has been designed to work with the existing IPv6 architectures. This 
solution shares the same drawback mentioned for ILNP, i.e.~when the two 
end-points move simultaneously and direct updates get lost, a service 
unavailability can occur.

\subsection{Solutions Between the Network and Transport Layers}

Approaches have been presented in literature that insert an intermediate 
layer between the network and transport layers of the protocol stack. This layer 
works on all the end-nodes involved in the communication, i.e., MN and CN (in a one-to-one 
communication). 
Notable examples in this class of solutions are the Host Identity Protocol (HIP) 
\cite{BokorZNJ09,rfc4423}, Hi3 \cite{GurtovKLN08}, Location Independent Addressing 
for IPv6 (LIN6) \cite{draft-teraoka-ipng,kunishi2000lin6}, MILSA, 
\cite{Pan:2010} and Level 3 Multihoming Shim Protocol for IPv6 (Shim6) 
\cite{rfc5533}. 
Based on these solutions, the location registry is a DNS-like mapping function 
that operates as a service outside the access networks and associates host 
identifiers to host locations. 

A limitation of these approaches is the requirement to modify the protocol stack in all the end-nodes involved in a communication. While it is reasonable to require that a MN installs additional software to support its mobility, the CN can be a fixed node that may not be interested in supporting the mobility of the MN.

\subsubsection{HIP, Hi3}
In the Host Identity Protocol (HIP), IP addresses are locators used for 
packets forwarding only, while the concept of host identity is employed as the 
identifier of a node \cite{BokorZNJ09,rfc4423}. The protocol introduces an 
``UPDATE'' message that can be sent by a MN making a handover to its CN. As 
mentioned for other approaches, this mechanism fails when both end points are 
mobile and perform a handover simultaneously. To overcome this limitation, a 
rendezvous server is exploited that can be queried to map identifiers to their 
related locators.

The Host Identity Indirection Infrastructure (Hi3) is a networking
architecture for mobile nodes, derived from the i3 and the Host Identity Protocol (HIP) \cite{GurtovKLN08,nikander2004host}. The basic idea is to allow an IP-based communication while using an indirection infrastructure to route the HIP control messages.
The advantages of using i3 as a control plane for HIP in Hi3 include protection from Denial of Service attacks, support for simultaneous mobility, and providing an initial rendezvous service.
However, in order for Hi3 to be effective, i3 proxies are required in the network.

\subsubsection{RANGI}
Similarly to HIP, the Routing Architecture for the Next 
Generation Internet (RANGI) separates location and 
identifier, adopting an ID/Locator
mapping system \cite{rangi}.
In RANGI host locators are ordinary IPv6 addresses and consist of two parts: 
the first part represents the locator domain, such as the organization 
affiliation, while the second part is the local host identifier, which is 
represented as an IPv4 address (included within the IPv6 address).
The mapping system is implemented as a distributed hash table.
A node can have multiple locators at the same time with the same identifier 
(i.e.~the IPv4 address) thus enabling multihoming. 
In this last case, border routers are responsible for modifying the locators, thus enabling multipath delivery strategies. As to mobility, no detailed solutions have been proposed. Similarly to HIP, it is suggested that some update messages can be exploited to inform the CN of handovers or rehoming.

\subsubsection{Shim6}
In Shim6, locators are IPv6 addresses, while an upper layer 
identifier (which is described also as an IPv6 address) is employed to identify 
end points \cite{rfc5533}. As in other approaches, a layer is added in between 
the network and transport layers.
Communication between two end points is performed through a 4-way handshake 
that allows nodes to determine the locators and upper-level identifiers.
To enable mobility, update messages are exploited to change the locators, as in 
HIP.
In case of outage, Shim6 
performs an end-to-end exploration of the available addresses using the Reachability Protocol (REAP) \cite{rfc5534},
and then
updates the locators. This approach is not suitable for highly dynamic 
environments as it is related to timer expiration and not to movement 
detection. 

\subsubsection{MILSA}
MILSA employs the sublayer (added on top of the network layer) to map identifiers exploited at the application layer with
network-level locators \cite{Pan:2010}. 
Mobility and multihoming might be supported, since the mapper makes changes to the MN’s locator which are transparent to the applications. However, some kind of global manager is required for node identifiers. Moreover, no algorithm for the management of handovers is specified.

\subsubsection{NIIA}
Node Identity Inter-networking Architecture (NIIA) employs a 
node identity layer between the network and transport layers that is exploited 
to perform routing \cite{niia,Schutz:2010}.
The architecture defines two basic components: locator domains and node identity 
routers. A locator domain is the abstraction of some kind of local network, 
having a consistent internal addressing and routing system. 
Node identity routers perform routing between nodes in different domains. Each node registers to one node identity router. Then, node identity routers are in charge of modifying the headers entering/leaving the local domain, changing the locators so that inter-domain routing is made possible.
Moreover, they act as proxies, hence enabling node mobility and multihoming. When a node changes its 
local domain, a registration phase is executed, and the node identity 
router of the original local domain of the MN is informed. Hereinafter, that 
router will act as a proxy for the MN. 
In the designed architecture, local domains can be organized in a hierarchical manner. Thus, inter-domain messages can pass through a sequence of local domains; this can introduce additional latencies.

\subsection{Solutions at the Transport Layer}

As for previous approaches, the use of solutions working at the transport-layer may require 
modifications of the applications on both the MN and CN to invoke the services 
offered by these solutions. 
An overview of the existing proposals follows.

\subsubsection{MN as proactive location registry}
The common approach of the protocols working at the 
transport-layer, such as the datagram-oriented Datagram Congestion Control 
Protocol (DCCP) \cite{rfc4340}, the stream-oriented Mobile Stream Control 
Transport Protocol (m-SCTP) \cite{Budzisz:2012}, the TCP enhancements 
TCP-migrate \cite{snoeren2001reconsidering} and MultiPath TCP (MPTCP) 
\cite{mptcp,Paasch:2012}, is as follows. 
Each given end-system plays the 
role of a proactive location registry that directly informs the CN whenever its 
configuration changes. 
Unfortunately, this approach may fail when both the end-systems are mobile. Indeed, if they change their IP configuration simultaneously, leaving their current network access point, they would become mutually unreachable. 
The mentioned schemes support multihoming. While DCCP was originally designed without support for multihoming, some improvements have been proposed to add this feature \cite{DCCPMOB}.

\subsubsection{MSOCKS, I-TCP}
MSOCKS uses an external proxy that performs TCP connection redirection \cite{msocks}. 
Such an external proxy is employed to split the end-to-end communication into 
two communications: namely MN-proxy and proxy-CN. This solution is referred to as TCP Splice. Hence, MSOCKS 
uses TCP Splice to migrate a connection when the MN changes its IP address (and 
potentially when it changes network interface). Since the connection between its 
CN and the proxy remains unchanged, the connection will not be interrupted and 
the CN will not be aware of the mobility. TCP packets are modified by the 
external proxy to create the illusion of a single, direct TCP connection between 
the end-points. 

Another example of this kind of approach is I-TCP, that splits the communication into two paths, using an external relay to route messages \cite{itcp}.

\subsubsection{ECCP}
The End-to-end Connection Control Protocol (ECCP) is an end-to-end approach 
that modifies the transport protocol into two sublayers, i.e.~i) a connection 
control, which manages connections, their constituent flows, and their
associated addresses, and ii) a data delivery functionality that is responsible for the typical management of the transport flow once a connection is established, like congestion control, flow control, reliability \cite{eccp}.
Thus, functionalities are introduced so that once a node wants to change its NIC, it informs the other node.
During the communication, end-points can add flows to an existing connection
in order to spread traffic over multiple NICs.

We mentioned that no end-to-end signaling
protocol can, by itself, handle the case when both hosts
reconfigure or change simultaneously the NIC in use.
To sort out this problem, the authors propose the use of a lightweight redirection cache in the local network
of either communicating hosts. This cache should keep short-lived
redirection state pointing to the new locations of hosts that
have recently migrated out of its network. When a MN moves, it sends a message to the
redirection box of its old network to add a pointer to its new
location. This additional scheme requires the modification of the network infrastructure and violates the end-to-end philosophy of the protocol.

\subsection{Solutions at the Session Layer}

\subsubsection{SIP-based approaches}
Several proposals exist that employ SIP to control the session of 
a multi-homed communication. For instance, the Terminal Mobility Support 
Protocol (TMSP) \cite{LimYLL09} exploits an auxiliary SIP server, as location 
registry placed outside the access networks, that maps a user identifier (e.g.~vittorio.ghini@unibo.it) to the current user location (i.e.~the IP address of his/her  
MN). 
Each MN has a SIP user agent that sends REGISTER messages to the SIP server in 
order to update its current location. INVITE messages are sent to establish 
communications with other nodes. 

Similarly, \cite{Udugama:2007} presents an architecture capable of managing vertical 
handoffs, by using a SIP-based approach. The scheme complies to the IP 
Multimedia Subsystem (IMS), a standardized overlay architecture for session 
control, authentication, authorization and accountability in all-IP networks.  
Another related proposal is the one presented in \cite{Kalmanek:2006} that only 
supports vertical handoffs from 3G networks to a WiFi network.

Session-layer solutions might not be efficient as they invoke an external 
localization service when an IP address reconfiguration occurs. In particular, 
the SIP-based services introduce an additional delay due to their 
message/response behavior; in case of reconfiguration, the MN interrupts the 
communication, sends a SIP signaling message to the CN and waits for the 
response before resuming the transmission.  

With this in view, the IHMAS work presented in \cite{Bellavista:2010} 
minimizes handoff delays by exploiting a SIP-based, IMS compliant, proactive 
mechanism that performs registration and renegotiation phases for new 
connections, while keeping the media flows active over old connections, if these 
are available. 

\subsubsection{Coping with NAT and Firewall Systems}

An example of a system that takes into account the possible presence of firewall and NAT systems is MMUSE \cite{SalsanoPMNV08}. 
This system requires an auxiliary SIP server (namely, the Session Border Controller, SBC) to be located at the edge of the autonomous system where the MN is entering. 
This autonomous system may be composed of several subnets using heterogeneous network technologies. While the MN moves across the subnets, each subnet provides the MN with a different IP address. 
The SBC combines the functionalities of SIP and RTP proxies, firewall and NAT systems, and intercepts the communications that enter and leave the network, in particular the SIP messages between the MN and its CN outside the network edge. 
The SBC sets up the firewall rules that allow the subsequent SIP, RTP and RTCP communications between
MN and CN based on the outgoing SIP messages. Moreover, when the MN moves to a different subnet and changes its IP address, the SBC modifies the outgoing datagram, in order to hide the current location of the MN from the CN.

The main limitation of MMUSE is that the MN traffic always needs to flow through a given SBC that resides in the edge of the network. 
This implies that the MN may only move inside a given autonomous system, but it cannot move across different networks administered by different organizations.

\subsection{Zero impact on Existing Network Infrastructures}

To limit the need for modifying the network infrastructure, the \textbf{external relay/proxy solutions} deploy the localization service at an external proxy, independent from both home and access networks, with no impact on the network infrastructures. 
This external proxy also incorporates the functionality of data relay to overcome NATs/firewalls. 
This class of solutions requires some modifications to the MN. Moreover, it splits the communications in two consecutive paths, from the MN to the proxy and from the proxy to the CN (as in all other solutions employing an additional relay to overcome the presence of firewalls).

The external relay/proxy class of host mobility solutions may be divided into 
two sub-classes: the \textbf{visible} class supports only those applications and 
protocols (such as Web and VoIP) that define the concept of proxy. It operates 
using a pair of explicit proxy software components, the one placed inside the MN and the
another on the relay/proxy external server.  
The application at the MN only needs to be configured to use the proxy running 
on the local host. 

On the other hand, the \textbf{invisible} class groups those solutions that operate as a tunnel between the MN and the external proxy; thus, applications running in the MN are not aware of these software entities. 
This sub-class does not depend on the application, but handles IP datagrams in different ways, based on the transport protocols. 
Similarly to the visible class, local proxy (at the MN) and external proxy software components are used. However, in this case, the application in the MN is unaware of the two proxies; in fact, the CN recognizes the external proxy server as its correspondent node
(i.e.,~the MN). 

\subsubsection{Visible Relay/Proxy Services and Early Packet Loss Detection}
These approaches exploit some additional software components on the network 
that act as proxy and/or relay. Since proxies need to manage application level 
data, these solutions are application-dependent. Thus, each system supports some 
specific application or, stated differently, each application protocol requires 
a specific implementation of the proxy that manages and routes application data.

\begin{figure}[h]
   \centering
   \includegraphics[width=\linewidth]{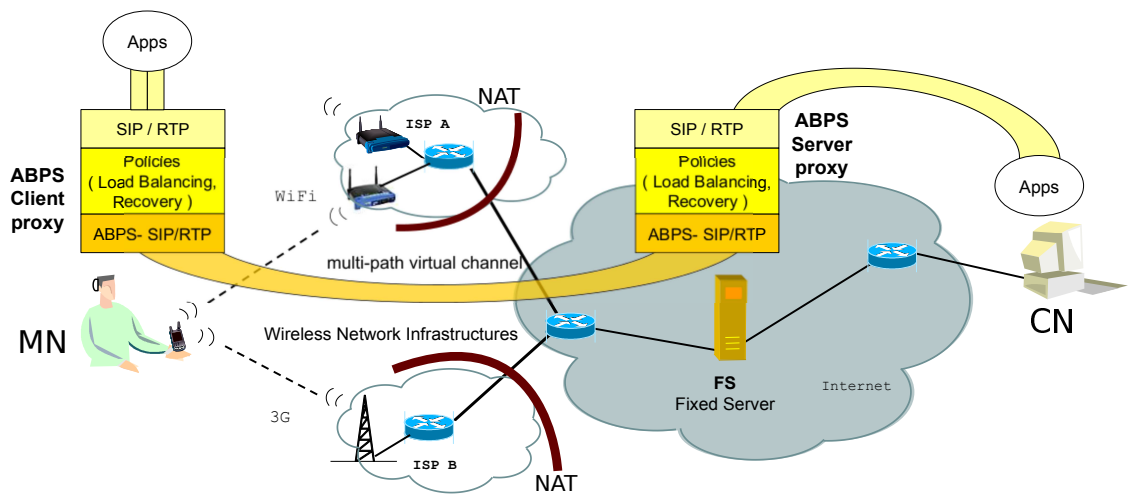}
   \caption{The ABPS-SIP/RTP architecture}
   \label{fig:abps}
\end{figure}

The highest-developed system within the visible relay class is the Always Best Packet Switching (ABPS-SIP/RTP) architecture, designed to support applications based on SIP/RTP such as VoIP and Video on Demand (VoD) \cite{GhiniJSS} (Figure \ref{fig:abps}). 
In contrast with the previously presented approaches, ABPS-SIP/RTP (or simply, ABPS) enables the transmission of each datagram through the most appropriate NIC among those which are active at the MN. 
The ABPS-SIP/RTP architecture operates at the session-layer using a so called ``proxy client'' on the MN and an auxiliary ``proxy server'' in an external server, allowing the MN to move across different autonomous systems. 
(The proxy acts as a relay as well, thus overcoming problems due to the presence of NATs and firewalls.)
A cross-layer technique is employed to monitor all the concurrent NICs 
which are available, their performances and those that become active (or inactive); 
based on their current status, an automatic reconfiguration is performed at the 
MN. This is accomplished by exploiting an implementation of the previously 
mentioned TED software module (see Section \ref{sec:tool_low_levels}).

This solution uses SIP-compliant proxy servers which interact with additional software modules, installed on the proxies working between the network and the transport layer; these modules are responsible for managing the application-layer data flows, enabling their transmission through different NICs. 
In practice, an ABPS proxy decides on a per-packet basis which is the best NIC to use to transmit the data. Moreover, each proxy adds a digital signature in the packet so that the proxy server may identify the sender, in spite of its possible different IP addresses. 
The use of such a signature technique transparently avoids the typical delays introduced by the two way message/response handshake of the SIP signaling phase.

It is important to note that all the approaches we introduced earlier usually consider the problem of changing a NIC in use as soon as it becomes unavailable. 
Thus, the decision to change communication technology is not taken based on some particular QoS metric; rather, it is taken based on the failure of the currently adopted NIC. 
In contrast, ABPS takes into consideration QoS metrics to identify the best NIC to use at any given moment. 
In particular, a cross-layer mechanism 
provides the applications with information about the successful (or unsuccessful) datagram transmissions through a given wireless access point. 

In essence, the advantages provided by the ABPS-SIP/RTP system are the following:
\begin{itemize}
 \item it works perfectly over all IP based networks and does not require any modification of the current network infrastructures;
 \item it is SIP compliant;
 \item it avoids delays occurring in classic SIP-based approaches when the MN changes its preferred NIC or its configuration. Indeed, in this case other schemes employ reconfiguration phases based on the exchange of INVITE messages, while the ABPS-SIP/RTP approach avoids this additional message exchange;
 \item it supports RTP-based applications such as VoIP and VoD services;
 \item it can cope with vertical handoffs, without introducing any additional delay during the passage from the use of a NIC to the next one since, as soon as this becomes available, it is promptly configured to work;
 \item it optimizes the use of NICs by deciding on a per-packet basis which is the best interface to be used, based on the monitored performances of the available networks and on the employed QoS metrics;
 \item it overcomes the presence of NATs and firewalls.
\end{itemize}
The principal limitation of this solution is that the ABPS-SIP/RTP architecture is strictly dedicated to SIP/RTP-based applications and cannot be exploited for other applications. 

Another approach that exploits an external proxy-based distributed approach to support HTTP-based Web 2.0 services has been presented in \cite{FerrettiG09}. In Figure \ref{fig:architectures} it is referred as CLW2A (Cross Layer architecture for Web 2.0 Applications).

These approaches are considered as working on a per-packet basis. Indeed, they 
work at a level of granularity which is finer than channel flows. However, they 
do not decide which NIC to use for each packet, since they schedule the NIC to 
use information that is available on top of the transport layer.
But they can easily switch from one NIC to another without wasting time in 
network and system reconfigurations. 

\subsubsection{Invisible Relay/Proxy Services}

The visible relay/proxy service solutions are not suitable for applications that are not designed to work with proxies, or more generally for legacy applications. 
To overcome this limitation, the invisible relay/proxy service class operates 
transparently from the application standpoint by intercepting messages sent by 
the applications in the MN and redirecting them to a local proxy.  
The local proxy delivers each message to the external relay/proxy using all the 
available NICs of the MN. In the MN, a virtual Ethernet interface is used 
(i.e. an Ethernet interface not linked to a physical one), and a particular 
routing table configuration causes the applications to bind their outgoing 
connections to that virtual interface.  

The earliest representative of this class was the ``hidden proxy'' 
\cite{ghi06}, developed on Linux systems and based on the iptables/netfilter 
mechanism and on the tun/tap virtual interface. 
This proxy was dedicated to TCP-based applications \cite{ghi06}. More 
recently, a similar architecture, called Universal Per-application Mobility
management using Tunnels (UPMT),
extends to the above mentioned hidden proxy to provide 
support for UDP-based applications \cite{Bonola:2009}. Both these 
architectures consider the problem of changing a NIC in use as soon as it 
becomes unavailable. 

Actually, it is conceivable that some solution can be devised that combines the 
use of an invisible external proxy with the features of early packet loss 
detection (in Figure \ref{fig:architectures} and in the following discussion 
we refer to such a kind of approach as Fast Reactive Hidden Proxy -- FRHP).
While at the time of writing, there are no available implementations of this 
strategy to support mobility management, this approach has been utilized in 
other contexts, such as vehicular networks \cite{Giordano:2012}.

\section{Comparison among Host Mobility Architectures}
\label{sec:comparison}

Tables \ref{tab}--\ref{tab2} summarize the discussion presented in this paper 
by providing a concise comparison of the different host mobility 
architectures we have presented.  
In both tables, each column refers to one system (or more systems 
with similar properties). 
Systems are organized based on the classes identified in the previous 
sections. Thus, we have pure end-to-end systems, hybrid end-to-end systems, 
systems employing some software entity placed within the MN's home network, 
systems that modify the access network, those employing some invisible external 
relay and, finally, those employing a visible external relay.

This qualitative comparison takes into account different criteria: Table 
\ref{tab} focuses on deployability, systems requirements, and issues concerned 
with the need to modify the applications, nodes or the 
network. 
Table \ref{tab2} focuses on performance issues. Rows display 
the criteria or features that each solution may have.
At a first glance, it can be observed that systems of the same class have 
similar properties.
At the time of writing, there are no available real implementations for some approaches, whereas for others, the related papers/documentation do not provide enough details and do not allow us to retrieve complete information on some performance parameters. 
Systems for which we do not have enough information are not ranked according to those specific performance metrics, and are instead marked with an “U” (i.e., it is unclear how the system would work in certain conditions).

Tables \ref{tab}--\ref{tab2} are described in detail as follows.
Table \ref{tab} is structured in three principal subtables named ``General'', ``Requirements'' and ``Need Modifications''. 
Table \ref{tab2} is structured in a single subtable labeled ``Performance''.
These subtables are described below individually.

\begin{landscape}
\begin{table*}[t]
\caption{Comparison among host mobility architectures: Requirements and general issues on the deployability}
\label{tab}
 \centering
 \scriptsize
 \begin{tabular}{|c|c|c|c|c|c|c|c|c|c|c|c|c|c|c|c|c|c|c|c|c|c|c|c|c|c|}
 \hline
 \hline
  & \multirow{2}{*}{Classes} 
    & \multicolumn{2}{c|}{Pure} & \multicolumn{6}{c|}{Hybrid} & 
    \multicolumn{3}{c|}{Home} & \multicolumn{9}{c|}{Access}
    & \multicolumn{3}{c|}{Invisible} & Visible\\
  & & \multicolumn{2}{c|}{E2E} & \multicolumn{6}{c|}{E2E} & 
    \multicolumn{3}{c|}{Network} & \multicolumn{9}{c|}{Network}
    & \multicolumn{3}{c|}{Proxy} & Proxy \\
 \hline
 \hline
    & \multirow{16}{*}{Criteria} &
\multirow{15}{*}{\rotatebox[origin=c]{-90}{TCP-migrate 
      \cite{snoeren2001reconsidering} -- MPTCP \cite{mptcp}}} &
      \multirow{16}{*}{\rotatebox[origin=c]{-90}{DCCP \cite{rfc4340} -- ECCP 
	  \cite{eccp} -- m-SCTP \cite{Budzisz:2012}}} & 
      \multirow{16}{*}{\rotatebox[origin=c]{-90}{HIP \cite{BokorZNJ09,rfc4423} 
	  -- Hi3 \cite{GurtovKLN08}}} & 
      \multirow{16}{*}{\rotatebox[origin=c]{-90}{LIN6 \cite{draft-teraoka-ipng} 
	  }} & 
      \multirow{16}{*}{\rotatebox[origin=c]{-90}{MSOCKS \cite{msocks} -- 
	  I-TCP \cite{itcp}}} & 
      \multirow{16}{*}{\rotatebox[origin=c]{-90}{Shim6 \cite{rfc5533}}} & 
      \multirow{16}{*}{\rotatebox[origin=c]{-90}{TMSP \cite{LimYLL09}, 
\cite{Udugama:2007}, \cite{Kalmanek:2006}, IHMAS \cite{Bellavista:2010}}} & 
      \multirow{16}{*}{\rotatebox[origin=c]{-90}{HIMA \cite{Shenoy:2005}}} & 
      \multirow{16}{*}{\rotatebox[origin=c]{-90}{MIPv6 \cite{Li:2009}, 
\cite{VanHanh:2008}, \cite{rfc-4140}, \cite{rfc5213}, \cite{Zhou:2010}, 
\cite{He:2010} }}      & 
      \multirow{16}{*}{\rotatebox[origin=c]{-90}{Monami6 \cite{rfc4861} -- 
		      Nemo \cite{Perera:2004}}} & 
      \multirow{16}{*}{\rotatebox[origin=c]{-90}{FlowMob \cite{Toseef:2008}}} & 
      \multirow{16}{*}{\rotatebox[origin=c]{-90}{SIP-IAPP \cite{WuYH07}}} & 
      \multirow{16}{*}{\rotatebox[origin=c]{-90}{LISP \cite{lisp}}} & 
      \multirow{16}{*}{\rotatebox[origin=c]{-90}{ILNP \cite{rfc6740}}} &
      \multirow{16}{*}{\rotatebox[origin=c]{-90}{GLI-Split \cite{gli-split}}} & 
      \multirow{16}{*}{\rotatebox[origin=c]{-90}{MILSA \cite{Pan:2010}}} & 
      \multirow{16}{*}{\rotatebox[origin=c]{-90}{RANGI \cite{rangi}}} & 
      \multirow{16}{*}{\rotatebox[origin=c]{-90}{NIIA
			\cite{niia,Schutz:2010}}} & 
      \multirow{16}{*}{\rotatebox[origin=c]{-90}{MMUSE \cite{SalsanoPMNV08}}} & 
      \multirow{16}{*}{\rotatebox[origin=c]{-90}{ROAM \cite{Zhuang:2005}}} & 
      \multirow{16}{*}{\rotatebox[origin=c]{-90}{Hidden Proxy \cite{ghi06}}} & 
      \multirow{16}{*}{\rotatebox[origin=c]{-90}{UPMT \cite{Bonola:2009}}} & 
      \multirow{16}{*}{\rotatebox[origin=c]{-90}{FRHP \cite{Giordano:2012}}} & 
      \multirow{16}{*}{\rotatebox[origin=c]{-90}{ABPS \cite{GhiniJSS} -- 
		      CLW2A \cite{FerrettiG09}}} \\
 & & & & & & & & & & & & & & & & & & & & & & & & &\\
 & & & & & & & & & & & & & & & & & & & & & & & & &\\
 & & & & & & & & & & & & & & & & & & & & & & & & &\\
 & & & & & & & & & & & & & & & & & & & & & & & & &\\
 & & & & & & & & & & & & & & & & & & & & & & & & &\\
 & & & & & & & & & & & & & & & & & & & & & & & & &\\
 & & & & & & & & & & & & & & & & & & & & & & & & &\\
 & & & & & & & & & & & & & & & & & & & & & & & & &\\
 & & & & & & & & & & & & & & & & & & & & & & & & &\\
 & & & & & & & & & & & & & & & & & & & & & & & & &\\
 & & & & & & & & & & & & & & & & & & & & & & & & &\\
 & & & & & & & & & & & & & & & & & & & & & & & & &\\
 & & & & & & & & & & & & & & & & & & & & & & & & &\\
 & & & & & & & & & & & & & & & & & & & & & & & & &\\
 & & & & & & & & & & & & & & & & & & & & & & & & &\\
\hline
\hline
  \multirow{4}{*}{\rotatebox[origin=c]{-90}{General}} & 
	\multirow{2}{*}{Deployability} & \multirow{2}{*}{Y} & \multirow{2}{*}{Y} 
& \multirow{2}{*}{Y} & & \multirow{2}{*}{Y} & & \multirow{2}{*}{Y} & 
\multirow{2}{*}{Y} & 
		& & & \multirow{2}{*}{Y} & & & & & & & \multirow{2}{*}{Y} & 
\multirow{2}{*}{Y} & \multirow{2}{*}{Y} & \multirow{2}{*}{Y} & 
\multirow{2}{*}{Y} & \multirow{2}{*}{Y} \\
    & & & & & & & & & & & & & & & & & & & & & & & & &\\
  \hhline{~-------------------------}
  & \multirow{2}{*}{Implementation} & \multirow{2}{*}{Y} & \multirow{2}{*}{Y} & 
\multirow{2}{*}{Y} & \multirow{2}{*}{Y} & \multirow{2}{*}{Y} & 
\multirow{2}{*}{Y} & \multirow{2}{*}{Y} &   & \multirow{2}{*}{Y} 
		& \multirow{2}{*}{Y} & \multirow{2}{*}{Y} & \multirow{2}{*}{Y} & 
\multirow{2}{*}{Y} & \multirow{2}{*}{Y} & \multirow{2}{*}{Y} &   &   & 
\multirow{2}{*}{Y} & \multirow{2}{*}{Y} &   & \multirow{2}{*}{Y} & 
\multirow{2}{*}{Y} &  & \multirow{2}{*}{Y} \\
    & & & & & & & & & & & & & & & & & & & & & & & & & \\
\hline
  \multirow{6}{*}{\rotatebox[origin=c]{-90}{Requirements}} & 
	Requires & & &  & \multirow{2}{*}{Y} & & \multirow{2}{*}{Y} & & & 
		\multirow{2}{*}{Y} &  \multirow{2}{*}{Y} & 
		\multirow{2}{*}{Y} & & \multirow{2}{*}{Y} & \multirow{2}{*}{Y} & \multirow{2}{*}{Y}
		 & \multirow{2}{*}{Y} & \multirow{2}{*}{Y} & \multirow{2}{*}{Y} & & & & & & \\
    &    IPv6 & & & & & & & & & & & & & & & & & & & & & & & &\\
  \hhline{~-------------------------}
    &    Requires & & &  & & & & & & \multirow{2}{*}{Y} & \multirow{2}{*}{Y} & \multirow{2}{*}{Y} & & & & & & & & & & & & &\\
    &    MIPv6 & & & & & & & & & & & & & & & & & & & & & & & &\\
  \hhline{~-------------------------}
    &   Requires & & &  & & & & & & \multirow{2}{*}{Y} & \multirow{2}{*}{Y} & \multirow{2}{*}{Y} & 
		 & & & &  \multirow{2}{*}{Y} & & & & & \multirow{2}{*}{Y} & & & \\
    &   IPSec & & & & & & & & & & & & & & & & & & & & & & & &\\
\hline
  \multirow{14}{*}{\rotatebox[origin=c]{-90}{Needs Modifications}} & 
      Protocol Stack & \multirow{2}{*}{Y} & \multirow{2}{*}{Y} & \multirow{2}{*}{Y}&  
              \multirow{2}{*}{Y} & \multirow{2}{*}{Y} & \multirow{2}{*}{Y} & \multirow{2}{*}{Y} & \multirow{2}{*}{Y} & 
	      \multirow{2}{*}{Y} & \multirow{2}{*}{Y} & \multirow{2}{*}{Y}  & & \multirow{2}{*}{Y} & \multirow{2}{*}{Y} & 
	      \multirow{2}{*}{Y} & \multirow{2}{*}{Y} & \multirow{2}{*}{Y} &
	      \multirow{2}{*}{Y} & \multirow{2}{*}{Y} & \multirow{2}{*}{Y} & \multirow{2}{*}{Y}
	       & \multirow{2}{*}{Y} & \multirow{2}{*}{Y} & \multirow{2}{*}{Y}\\
  &   in MN & & & & & & & & & & & & & & & & & & & & & & & & \\
  \hhline{~-------------------------}
  &  Protocol Stack & \multirow{2}{*}{Y} & \multirow{2}{*}{Y}&  
              \multirow{2}{*}{Y} & \multirow{2}{*}{Y} & \multirow{2}{*}{Y} & 
              \multirow{2}{*}{Y} &  \multirow{2}{*}{Y} &  \multirow{2}{*}{Y} & 
              \multirow{2}{*}{Y} & \multirow{2}{*}{Y} & \multirow{2}{*}{Y} & & &  \multirow{2}{*}{Y} &
               \multirow{2}{*}{Y} & \multirow{2}{*}{Y} & \multirow{2}{*}{Y} &
                \multirow{2}{*}{Y} & & \multirow{2}{*}{Y} & 
	      & & & \\
  &  in CN & & & & & & & & & & & & & & & & & & & & & & & &\\
  \hhline{~-------------------------}
  &  Applications & &  \multirow{2}{*}{Y} & & & & &
	      \multirow{2}{*}{Y} & \multirow{2}{*}{Y} & & & & & & \multirow{2}{*}{Y} & \multirow{2}{*}{Y} & 
	      &  \multirow{2}{*}{Y} & \multirow{2}{*}{Y} &  & & 
	      & &  & \\
  &  in MN & & & & & & & & & & & & & & & & & & &  & & & & & \\
  \hhline{~-------------------------}
  &  Applications &  & \multirow{2}{*}{Y} & & & & &
	      \multirow{2}{*}{Y} & \multirow{2}{*}{Y} & & &  & & & \multirow{2}{*}{Y} & \multirow{2}{*}{Y} & 
	      &  \multirow{2}{*}{Y} & \multirow{2}{*}{Y} & & 
	       & & & & \\
  &  in CN & & & & & & & & & & & & & & & & & & & & & & & & \\
  \hhline{~-------------------------}
  &  Access & & & &  & & & & & \multirow{2}{*}{Y} & 
              \multirow{2}{*}{Y}  & & \multirow{2}{*}{Y}  & &  \multirow{2}{*}{Y}  & \multirow{2}{*}{Y}  & \multirow{2}{*}{Y}  & \multirow{2}{*}{Y}  & \multirow{2}{*}{Y}  & & & & & &\\
  &  Networks & & & & & & &  & & & & & & & & & & & & & & & & &\\
  \hhline{~-------------------------}
  &  Border   & & & &  & & & & & & & & \multirow{2}{*}{Y} & \multirow{2}{*}{Y} & \multirow{2}{*}{Y} 
              & \multirow{2}{*}{Y} & \multirow{2}{*}{Y} & \multirow{2}{*}{Y} & \multirow{2}{*}{Y}
              & \multirow{2}{*}{Y} & & & & & \\
  &  Gateways & & & & & & & & & & & & & & & & & & & & & & & &\\
  \hhline{~-------------------------}
  &  Requires    & & & & & \multirow{2}{*}{Y} &  & & & & & & & & & & & & & & \multirow{2}{*}{Y} 
	      & \multirow{2}{*}{Y} & 
	      \multirow{2}{*}{Y} & \multirow{2}{*}{Y} & \multirow{2}{*}{Y} \\
  &  External Relay/Proxy & & & & & & & & & & & & & & & & & & & & & & & &\\
\hline
\hline
\multicolumn{26}{|l|}{Symbols -- Y: yes}\\
\hline
\hline
\end{tabular}
\end{table*}
\end{landscape}

\begin{landscape}
\newgeometry{hmargin=0.5cm}
\begin{table*}[t]
\caption{Comparison among host mobility architectures: Performance}
\label{tab2}
 \centering
 \scriptsize
 \begin{tabular}{|c|c|c|c|c|c|c|c|c|c|c|c|c|c|c|c|c|c|c|c|c|c|c|c|c|c|}
 \hline
 \hline
  & \multirow{2}{*}{Classes} 
    & \multicolumn{2}{c|}{Pure} & \multicolumn{6}{c|}{Hybrid} & 
    \multicolumn{3}{c|}{Home} & \multicolumn{9}{c|}{Access}
    & \multicolumn{3}{c|}{Invisible} & Visible\\
  & & \multicolumn{2}{c|}{E2E} & \multicolumn{6}{c|}{E2E} & 
    \multicolumn{3}{c|}{Network} & \multicolumn{9}{c|}{Network}
    & \multicolumn{3}{c|}{External Proxy} & Proxy \\
 \hline
 \hline
    & \multirow{16}{*}{Criteria} &
\multirow{15}{*}{\rotatebox[origin=c]{-90}{TCP-migrate 
      \cite{snoeren2001reconsidering} -- MPTCP \cite{mptcp}}} &
      \multirow{16}{*}{\rotatebox[origin=c]{-90}{DCCP \cite{rfc4340} -- ECCP 
	  \cite{eccp} -- m-SCTP \cite{Budzisz:2012}}} & 
      \multirow{16}{*}{\rotatebox[origin=c]{-90}{HIP \cite{BokorZNJ09,rfc4423} 
	  -- Hi3 \cite{GurtovKLN08}}} & 
      \multirow{16}{*}{\rotatebox[origin=c]{-90}{LIN6 \cite{draft-teraoka-ipng} 
	  }} & 
      \multirow{16}{*}{\rotatebox[origin=c]{-90}{MSOCKS \cite{msocks} -- 
	  I-TCP \cite{itcp}}} & 
      \multirow{16}{*}{\rotatebox[origin=c]{-90}{Shim6 \cite{rfc5533}}} & 
      \multirow{16}{*}{\rotatebox[origin=c]{-90}{TMSP \cite{LimYLL09}, 
\cite{Udugama:2007}, \cite{Kalmanek:2006}, IHMAS \cite{Bellavista:2010}}} & 
      \multirow{16}{*}{\rotatebox[origin=c]{-90}{HIMA \cite{Shenoy:2005}}} & 
      \multirow{16}{*}{\rotatebox[origin=c]{-90}{MIPv6 \cite{Li:2009}, 
\cite{VanHanh:2008}, \cite{rfc-4140}, \cite{rfc5213}, \cite{Zhou:2010}, 
\cite{He:2010}}} & 
      \multirow{16}{*}{\rotatebox[origin=c]{-90}{Monami6 \cite{rfc4861} -- 
		      Nemo \cite{Perera:2004}}} & 
      \multirow{16}{*}{\rotatebox[origin=c]{-90}{FlowMob \cite{Toseef:2008}}} & 
      \multirow{16}{*}{\rotatebox[origin=c]{-90}{SIP-IAPP \cite{WuYH07}}} & 
      \multirow{16}{*}{\rotatebox[origin=c]{-90}{LISP \cite{lisp}}} & 
      \multirow{16}{*}{\rotatebox[origin=c]{-90}{ILNP \cite{rfc6740}}} &
      \multirow{16}{*}{\rotatebox[origin=c]{-90}{GLI-Split \cite{gli-split}}} & 
      \multirow{16}{*}{\rotatebox[origin=c]{-90}{MILSA \cite{Pan:2010}}} & 
      \multirow{16}{*}{\rotatebox[origin=c]{-90}{RANGI \cite{rangi}}} & 
      \multirow{16}{*}{\rotatebox[origin=c]{-90}{NIIA
			\cite{niia,Schutz:2010}}} & 
      \multirow{16}{*}{\rotatebox[origin=c]{-90}{MMUSE \cite{SalsanoPMNV08}}} & 
      \multirow{16}{*}{\rotatebox[origin=c]{-90}{ROAM \cite{Zhuang:2005}}} & 
      \multirow{16}{*}{\rotatebox[origin=c]{-90}{Hidden Proxy \cite{ghi06}}} & 
      \multirow{16}{*}{\rotatebox[origin=c]{-90}{UPMT \cite{Bonola:2009}}} & 
      \multirow{16}{*}{\rotatebox[origin=c]{-90}{FRHP \cite{Giordano:2012}}} & 
      \multirow{16}{*}{\rotatebox[origin=c]{-90}{ABPS \cite{GhiniJSS} -- 
		      CLW2A \cite{FerrettiG09}}} \\
 & & & & & & & & & & & & & & & & & & & & & & & & &\\
 & & & & & & & & & & & & & & & & & & & & & & & & &\\
 & & & & & & & & & & & & & & & & & & & & & & & & &\\
 & & & & & & & & & & & & & & & & & & & & & & & & &\\
 & & & & & & & & & & & & & & & & & & & & & & & & &\\
 & & & & & & & & & & & & & & & & & & & & & & & & &\\
 & & & & & & & & & & & & & & & & & & & & & & & & &\\
 & & & & & & & & & & & & & & & & & & & & & & & & &\\
 & & & & & & & & & & & & & & & & & & & & & & & & &\\
 & & & & & & & & & & & & & & & & & & & & & & & & &\\
 & & & & & & & & & & & & & & & & & & & & & & & & &\\
 & & & & & & & & & & & & & & & & & & & & & & & & &\\
 & & & & & & & & & & & & & & & & & & & & & & & & &\\
 & & & & & & & & & & & & & & & & & & & & & & & & &\\
 & & & & & & & & & & & & & & & & & & & & & & & & &\\
\hline
\hline
  \multirow{26}{*}{\rotatebox[origin=c]{-90}{Performance}} & 
    Support to &  & & \multirow{2}{*}{Y} & \multirow{2}{*}{Y} &  
	      & \multirow{2}{*}{Y} & 
	      \multirow{2}{*}{Y} &  \multirow{2}{*}{Y} & 
	      \multirow{2}{*}{Y} & \multirow{2}{*}{Y} & \multirow{2}{*}{Y} 
	      & \multirow{2}{*}{Y} & 
	      \multirow{2}{*}{Y} &  & & & & & \multirow{2}{*}{Y} 
	      & & & \multirow{2}{*}{E} & 
	      \multirow{2}{*}{Y} & \multirow{2}{*}{Y} \\
    & SIP/RTP & & & & & & & & &  & & & & & & & & & & & & & & &\\
  \hhline{~-------------------------}
    & Support to & \multirow{2}{*}{Y} &  & \multirow{2}{*}{Y} & \multirow{2}{*}{Y} & 
	      \multirow{2}{*}{Y}	&
	      \multirow{2}{*}{Y} & & \multirow{2}{*}{Y} & \multirow{2}{*}{Y} & 
	      \multirow{2}{*}{Y} & \multirow{2}{*}{Y} & 
	      \multirow{2}{*}{Y} & 
	      \multirow{2}{*}{Y} & & \multirow{2}{*}{Y} & \multirow{2}{*}{Y} & \multirow{2}{*}{Y} 
	      & \multirow{2}{*}{Y} & & & \multirow{2}{*}{Y}  & \multirow{2}{*}{Y} & 
	      \multirow{2}{*}{Y} & \\
    & TCP  & & & & & & & & & & & & & & & & & & & & & & & & \\
  \hhline{~-------------------------}
    & Support to  & & & \multirow{2}{*}{Y} & \multirow{2}{*}{Y} & &
	      \multirow{2}{*}{Y} & & \multirow{2}{*}{Y} & \multirow{2}{*}{Y} & 
	      \multirow{2}{*}{Y} & \multirow{2}{*}{Y} & \multirow{2}{*}{Y} & 
	      \multirow{2}{*}{Y} & &  \multirow{2}{*}{Y} & \multirow{2}{*}{Y} &
	       \multirow{2}{*}{Y} & \multirow{2}{*}{Y} & & & & \multirow{2}{*}{Y} & 
	      \multirow{2}{*}{Y} & \\
    & UDP &  & & & & & & & & & & & & & & & & & &  & & & & &\\
  \hhline{~-------------------------}
    & Support to  & & & & & & & & \multirow{2}{*}{Y} & \multirow{2}{*}{Y} & 
	       \multirow{2}{*}{Y} & & \multirow{2}{*}{Y} & 
	      \multirow{2}{*}{Y} & & & \multirow{2}{*}{Y} & & & & & \multirow{2}{*}{Y} & \multirow{2}{*}{E} & 
	      \multirow{2}{*}{Y} & \\
    & Legacy App & & & & & & & & & & & & & & & & & & & & & & & & \\
  \hhline{~-------------------------}
    & Sender  & & & & & \multirow{2}{*}{Y} & \multirow{2}{*}{Y} & &  & \multirow{2}{*}{Y} & 
	      \multirow{2}{*}{Y} &  & \multirow{2}{*}{Y} & 
	      \multirow{2}{*}{Y} & \multirow{2}{*}{Y} &\multirow{2}{*}{Y} &\multirow{2}{*}{Y} &
	      \multirow{2}{*}{Y} &\multirow{2}{*}{Y} & & \multirow{2}{*}{Y} & & \multirow{2}{*}{Y} & 
	      \multirow{2}{*}{Y} & \multirow{2}{*}{Y} \\
    & Identification & & & & & & & & & & & & & & & & & & & & & & & &\\
  \hhline{~-------------------------}
    & Exploit simultaneously &  & & & & & & &  & &
	      \multirow{2}{*}{Y}  & \multirow{2}{*}{Y} & &
	      \multirow{2}{*}{Y} & \multirow{2}{*}{Y} & \multirow{2}{*}{Y} & \multirow{2}{*}{Y} & & \multirow{2}{*}{Y} & & & \multirow{2}{*}{Y} & \multirow{2}{*}{E} & 
	      \multirow{2}{*}{Y} & \multirow{2}{*}{Y} \\
    & multiple NICs & & & & &  & & & & & & & & & & & & & & & & & & &\\
  \hhline{~-------------------------}
    & \multirow{2}{*}{Granularity} & \multirow{2}{*}{C} & 
	      \multirow{2}{*}{C} & \multirow{2}{*}{N} & \multirow{2}{*}{N} & \multirow{2}{*}{N} & \multirow{2}{*}{N} & \multirow{2}{*}{N} & \multirow{2}{*}{N} &  \multirow{2}{*}{N} & \multirow{2}{*}{N} & \multirow{2}{*}{C} & 
	      \multirow{2}{*}{N} & \multirow{2}{*}{C} & \multirow{2}{*}{N} & \multirow{2}{*}{N} &
	      \multirow{2}{*}{N} & \multirow{2}{*}{N} & \multirow{2}{*}{N} & \multirow{2}{*}{C} & 
	      \multirow{2}{*}{N} & \multirow{2}{*}{P} & \multirow{2}{*}{P} & \multirow{2}{*}{P} & 
	      \multirow{2}{*}{P} \\
    & & & & & & & & & & & & & & & & & & & & & & & & &\\
  \hhline{~-------------------------}
    & Overcome & & & & & &   & & & & & & & 
	       & &  &  & & & \multirow{2}{*}{Y} & & \multirow{2}{*}{Y} & \multirow{2}{*}{Y} & 
	      \multirow{2}{*}{Y} & \multirow{2}{*}{Y} \\
    & FW and NAT & & & & &  & & & & & & & & & & & & & & & & & & &\\
  \hhline{~-------------------------}
    & Unavailability Interval &  M & M & \multirow{2}{*}{M} &
	      \multirow{2}{*}{M} & \multirow{2}{*}{M} & \multirow{2}{*}{M} & 
	      \multirow{2}{*}{H} & \multirow{2}{*}{H} & \multirow{2}{*}{H} & 
	      \multirow{2}{*}{H} &  \multirow{2}{*}{H} & \multirow{2}{*}{L} & 
	      M & M & M & \multirow{2}{*}{U} & \multirow{2}{*}{U} & \multirow{2}{*}{U} 
	      & H & \multirow{2}{*}{M} & \multirow{2}{*}{M} & \multirow{2}{*}{M} & 
	      \multirow{2}{*}{L} & \multirow{2}{*}{L} \\
    & Due to Handover & $\infty$ &  $\infty$ &  & & & & & & & & & & 
	      $\infty$ & $\infty$ & $\infty$ & & & & $\infty$ & & & & & \\
  \hhline{~-------------------------}
    & Continuity & \multirow{2}{*}{L} & \multirow{2}{*}{L} & \multirow{2}{*}{L} & 
	      \multirow{2}{*}{L} & \multirow{2}{*}{M} & \multirow{2}{*}{L} & 
	      \multirow{2}{*}{L} & \multirow{2}{*}{L} & \multirow{2}{*}{L} & 
	      \multirow{2}{*}{H} &  \multirow{2}{*}{H} 
	      & \multirow{2}{*}{L} & 
	      \multirow{2}{*}{M} & \multirow{2}{*}{M} & \multirow{2}{*}{M} & 
	      \multirow{2}{*}{U} & \multirow{2}{*}{U} & \multirow{2}{*}{U} & 
	      \multirow{2}{*}{M} & \multirow{2}{*}{M} & \multirow{2}{*}{H} & 
	      \multirow{2}{*}{H} & \multirow{2}{*}{H} & \multirow{2}{*}{H} \\
    & Interval & & & & & & & & &  & & & & & & & & & & & & & & &\\
  \hhline{~-------------------------}
    & E2E & \multirow{2}{*}{L}  & \multirow{2}{*}{L} & 
	      \multirow{2}{*}{L} & \multirow{2}{*}{L} & M 
	      & \multirow{2}{*}{L} & 
	      \multirow{2}{*}{L} & \multirow{2}{*}{L} & L & 
	      L  & \multirow{2}{*}{L} & M & \multirow{2}{*}{M} & 
	      L & \multirow{2}{*}{U} & \multirow{2}{*}{U} & \multirow{2}{*}{U} & 
	      \multirow{2}{*}{U} 
	      & M & M & M & M & M & M\\
    & Latency & &  & & & H & & & & H &  H & & H & & H & & & & & H & H & H & H & H & H\\
    & (best case) &  $\sharp$ & $\sharp$ & $\sharp$ & $\sharp$ & $\sharp\flat$ & $\sharp$ & 
	      $\sharp$ & $\sharp$ &  $\sharp\xi$ & $\sharp\xi$ & 
	      $\sharp$ & & $\sharp$ & $\sharp\xi$ & & & & &
	      $\flat$ & $\flat$ & $\flat$ & $\flat$ & $\flat$ & $\flat$\\
  \hhline{~-------------------------}
    & E2E Latency with & \multirow{3}{*}{H}  & \multirow{3}{*}{H} & 
	      \multirow{3}{*}{H} & \multirow{3}{*}{H} & M 
	      & \multirow{3}{*}{H} & 
	      \multirow{3}{*}{H} & \multirow{3}{*}{H} & V & 
	      V &  V & M & V &  \multirow{3}{*}{U} & \multirow{3}{*}{U} & \multirow{3}{*}{U} & \multirow{3}{*}{U} & \multirow{3}{*}{U} & M & M & M & M & M & M\\
    & Symmetric FW & & & &  & H & & & & H & H & H & H & H & & & & & & H & H & H & H & H & H\\
    & (common case) &  & &    &  & $\flat$ &  & &  & & & & & & & & & & & & & & $\flat$ & $\flat$ & $\flat$ \\
\hline
\hline
\multicolumn{26}{|l|}{Symbols -- Y: yes; E: can be extended to support it; C: channel; N: node; P: packet; L: low; M: medium; H: high; VH: very high; U: unclear}\\
\multicolumn{26}{|l|}{($\xi$) the E2E latency becomes high if the return routability fails due to firewall presence}\\
\multicolumn{26}{|l|}{($\sharp$) the E2E latency becomes high when it is necessary an external relay to overcome firewall}\\
\multicolumn{26}{|l|}{($\flat$) the E2E latency becomes high if the relay server is far from the MN}\\
\hline
\hline
\end{tabular}
\end{table*}
\restoregeometry
\end{landscape}

\subsection{General}

This subtable of Table \ref{tab} 
contains two criteria, i.e.,~``deployability'' and 
``implementation''. 

\subsubsection{Deployability}
The deployability criterion ranks systems that can be effectively deployed on the 
current Internet. This feature is derived from the characteristics analyzed in 
the rest of the table.

In substance, system architectures are considered as deployable when they work over 
IPv4 networks, and do not require modifications to the backbone. In general, 
pure end-to-end solutions and those that resort to an external 
proxy are deployable. 
Some hybrid end-to-end approaches are deployed (MSOCKS, I-TCP, TMSP, HIMA), as 
well as some proposals that modify the access 
network only (however, one might argue that this requirement prevents the 
deployability of the systems in the real world), i.e.~SIP-IAPP, \cite{WuYH07}, 
MMUSE, ROAM.

\subsubsection{Implementation} 
The implementation feature indicates whether or not some real prototype 
implementation is available. Implementations or descriptions of implementation can 
be found for the following systems: 
ABPS \cite{GhiniJSS};
DCCP \cite{dccp_imp};
FLowMob \cite{MeliaBOGC11};
Hi3 \cite{hi3};
Hidden Proxy \cite{ghi06};
HIP \cite{LundbergJanne-ols2003,openhip_tech};
I-TCP \cite{itcp};
LIN6 \cite{draft-teraoka-ipng,kunishi2000lin6};
LISP \cite{openlisp,lispmob};
MCoA (monami6) and NEMO \cite{nemo1,umip,tahi};
MIPv6 \cite{Li:2009,nemo1,umip,tahi};
MMUSE \cite{mmuse_code};
MPTCP \cite{mptcp,Paasch:2012};
MSOCKS \cite{msocks};
NIIA \cite{Schutz20101142};
Shim6 \cite{Barre:2011,openhip_tech};
SIP-IAPP \cite{802f02};
TCP-migrate \cite{tcp-migrate};
TMSP \cite{LimYLL09};
UPMT \cite{Bonola:2009,upmtRoma}.

Below, we provide a brief description of the implementation software 
projects, for the mentioned architectures, available at the time of writing.
The systems described are ordered alphabetically.

\subsubsection*{ABPS} 
A detailed description of the implementation of the ABPS system is 
reported in \cite{GhiniJSS}. The software is available upon direct request to 
the authors. Current versions of ABPS are available for Linux and Android 
systems.

\subsubsection*{DCCP}
DCCP-TP is an implementation of DCCP optimized for portability 
\cite{dccp_imp}. The site provides source code downloads and documentation.
Such implementation includes many DCCP features, including IPv4 NAT encapsulation.
It has been implemented in C language, for Linux systems, and released under 
the GNU Lesser \ac{GPL} v2.1.

\subsubsection*{FLowMob} 
In \cite{MeliaBOGC11}, an implementation for Linux operating systems of PMIP 
and its FlowMob extension was built to validate the proposal. 

\subsubsection*{Hi3} 
In \cite{hi3}, a prototype implementation of the Hi3 software architecture is 
described. It has been conceived as a layered agent based architecture to 
separate the different functionalities of the system. The main result of such a 
development process was the validation of the Hi3 model architecture.

\subsubsection*{Hidden Proxy} 
A prototype implementation for Linux systems of the Hidden Proxy is described 
in \cite{ghi06}. The aim of the implementation was to assess the viability of 
the proposal. The software is available upon direct request to the authors.

\subsubsection*{HIP} 
The OpenHIP project is developing free, open source software implementing the 
HIP protocol \cite{openhip_tech}. The aim is to develop client software for 
Linux, BSD, Mac OS X, and Windows operating systems, plus some tools for 
experimenting with HIP (e.g., network analyzers).

Another IPv6 based implementation of HIP for Linux 
systems is presented in \cite{LundbergJanne-ols2003}.

Finally, Hip4Inter.net was another effort for implementing the base HIP (whose 
Web site was unreachable at the moment of writing). FreeBSD and 
Linux were supported.

\subsubsection*{I-TCP} 
The paper presenting the I-TCP also provides a description of a prototype 
implementation \cite{itcp}. This implementation includes 
modifications to the TCP and Mobile-IP code on the routers and MNs. 
Authors state that no modifications are needed in the Unix kernel in the MN.
In the MN, the I-TCP library provides the \ac{API} for the I-TCP functions
which are similar to the socket related system 
calls in Unix. A further description of the implementation and a performance 
evaluation are reported in \cite{itcp:perf}.

\subsubsection*{LIN6} 
In \cite{draft-teraoka-ipng,kunishi2000lin6}, a prototype implementation of 
LIN6 on NetBSD/i386 is presented. The main goal of this 
effort was to validate the described system architecture.

\subsubsection*{LISP} 
OpenLISP is an open source implementation of the LISP Protocol running in the 
kernel of the FreeBSD Operating System \cite{openlisp}.
It implements a new type of sockets, called Mapping Sockets, providing an 
\ac{API} that can be used by any user space process.

LISPmob is another open-source LISP implementation for Linux, 
Android and OpenWrt \cite{lispmob}.

\subsubsection*{MIPv6, MCoA monami6 and NEMO} 
There are several available implementations of MIPv6 
\cite{Li:2009,nemo1,umip,tahi}. However, only some examples  taken from seminal articles will be presented here, together with their related projects.

UMIP is an open source implementation of Mobile IPv6 and NEMO Basic Support for 
Linux. It is released under the \ac{GPL} v2 \cite{umip}.

The Nautilus6 working group provided Linux and BSD reference implementations of 
IPv6 related libraries and IPv6 applications \cite{nemo1}. In particular, 
this working group developed SHISA (an implementation of Mobile IPv6 and 
NEMO), NEPL (a NEMO platform for Linux), and ATLANTIS (a NEMO Basic Support 
implementation for NetBSD).

The TAHI project was concluded at the end of 2012 by providing 
implementations of several protocols and test tools, e.g., IPv6 core, IPsec,
DHCPv6, MIPv6, NEMO \cite{tahi}.

\subsubsection*{MMUSE} 
An implementation of MMUSE is available in \cite{mmuse_code}. The prototype is 
written in Java language (which makes it a multi-platform solution) and uses mjsip as SIP 
stack. The source code is released under the \ac{GPL} v2.

\subsubsection*{MPTCP}
Implementations of MPTCP for Linux systems are described in 
\cite{mptcp,Paasch:2012}. The source code is available in \cite{mptcp}.
Three modes of operation are available, i.e.~i) the regular MPTCP, where 
all active NICs (and thus multiple TCP flows among two nodes) are used 
concurrently, ii) a backup mode, that uses only a subset of possible 
communication flows, iii) a single-path mode where a single sub flow is used; 
when the NIC goes down, a hard handover (break-before-make) is performed and a 
new TCP sub flow is created.

\subsubsection*{MSOCKS} 
The implementation of a MSOCKS library is described in \cite{msocks}. It is a 
shim library that sits between the application and the kernel on the mobile 
node. Its task is to provide an interface to the application, which is identical to that 
of the Berkeley Sockets \ac{API}, while internally using the normal TCP stack 
of the 
kernel to provide mobility functions. 
On Windows platforms, the MSOCKS library was designed to be a DLL that fits 
between the application and the WinSock DLL. Moreover, the paper mentions a BSD OS 
implementation, using the shared library support, but no further 
details are given.

\subsubsection*{NIIA} 
In \cite{Schutz20101142}, the authors describe a prototype implementation of NIIA, 
built based on the hip4inter.net code base.

\subsubsection*{Shim6} 
The OpenHIP project mentioned above also provides 
an implementation of Shim6 \cite{openhip_tech}. The idea was to 
provide an open-source Linux implementation of Shim6 that borrows from (and can 
later integrate with) the HIP implementation. This would lead to an eventual 
evolution towards a combined Shim6/HIP implementation.

LinShim6 (and its variant, MipShim6) is another project that implements Shim6 
\cite{Barre:2011}.

\subsubsection*{TCP-migrate} 
A Linux-based implementation of TCP-migrate has been released as free software, 
under the \ac{GPL} v2 \cite{tcp-migrate}. 

\subsubsection*{TMSP} 
In \cite{LimYLL09}, an implementation on Linux kernel of TMSP is described, 
followed by a performance evaluation. The system was tested with video 
streaming and video conferencing applications.  

\subsubsection*{UPMT} 
UPMT has been implemented for mainstream Linux and then ported on the 
Android OS \cite{Bonola:2009,upmt:website}. The source code is available under 
the \ac{GPL}. For demonstration purposes, a UPMT Linux Live 
distribution is provided where both UPMT mobile and anchor nodes can be set up.

\subsection{Requirements}

The requirements subtable describes what protocols and features are necessary for each given architecture.
In particular, we mark systems that require the presence of IPv6, MIPv6 or
IPSec. As shown in Table \ref{tab}, all solutions that exploit a software entity 
in the home network do require an IPv6 approach. Indeed, these solutions are 
designed to work with IPv6 or the like, and do exploit a \ac{HA}.

Some systems in the hybrid end-to-end class, i.e.~LIN6 and Shim6 (as their
names suggest), employ IPv6 as well. The other approaches, that have been devised to 
work over IPv6 nets, require a separation between the identifier 
and locator, i.e.~GLI-Split, ILNP, LISP, MILSA, NIIA, RANGI.

MIPv6 is needed by systems belonging to the Home Network class. These systems make use of IPsec, too. IPsec is used by MILSA and Hidden Proxy.

\subsection{Needs Modifications}

This subtable in Table \ref{tab} summarizes the changes that need to be introduced in the network entities of the examined architectures (e.g. infrastructures, protocols, terminals, applications) so that they can be deployed in real scenarios. 
These criteria are particularly important since they allow us to measure the applicability of a solution.
In this way, the requirement for modifying a given entity prevents the simple deployment of the solution and its practical use. 

The need for modifying the protocol stack at a MN (see the row ``Protocol Stack 
in MN'') is not a major limitation because the MN exploits the advantage of
mobility, which is a valid reason for users to install some additional 
software in their MN. Indeed, almost all the considered systems do require 
some particular changes to the MN, and if they do not, they still require 
the applications running on the MN to be modified (i.e.~TMSP \cite{LimYLL09}, 
HIMA \cite{Shenoy:2005}, see the row ``Applications in the MN''), or the 
access network to provide specific features and some border gateway to be 
present (i.e.~SIP-IAPP, \cite{WuYH07}, see rows ``Access Network'' and ``Border 
Gateways'').

Conversely, modifying the CN to ease the mobility of other nodes (i.e.~the MN) 
is not a convenient practice and it is very likely that those changes will never be deployed in real contexts. 
The list of systems that require such changes is reported in the row labeled ``Protocol Stack in CN''.
Basically, systems that do not have this requirement, assume the presence of an 
additional relay/proxy or a border gateway that manage packets to be sent to 
the CN.

Analogous considerations refer to those changes which are needed in the applications executed in the MN and in the CN (see rows “Applications in the MN” and “Applications in the CN”). 
Legacy applications cannot be modified. 
Owing to the above observations we can conclude that it would be difficult to see pure or hybrid end-to-end host mobility architectures being deployed in real, current scenarios. 
Moreover, we mentioned that systems in the access network class, that employ the identifier- locator distinction, are not designed for IPv4 networks.

Modifications to the network infrastructures require labour-intensive work, as well as reconfiguration and modification of the networks.
This is particularly true when a solution needs to modify all the access networks or border gateways. 
As mentioned, rows ``Access Network'' and ``Border Gateways'' report those systems that have such requirements.
They basically fall in the home network–based and access network–based host mobility classes.
Thus, the deployment of home network–based and access network–based host mobility architectures suffers from a massive inertia, due to the difficulties of deploying them in the existing Internet.

Solutions that rely on an external relay/proxy server are deployable 
because they do not impact on the existing infrastructures (see the ``Require 
External Relay'' row in the table).

The above considerations represent an important claim. We are indeed arguing 
that it is quite unlikely that the great majority of well known systems will be
employed in real scenarios, at least over a short or medium time period.

\subsection{Performance}

The Performance subtable in Table \ref{tab2} consists of criteria that describe the \ac{QoS} provided by the considered architectures. 
In particular, these criteria include: the ability of providing support 
to different applications and the protocols commonly employed to build them 
(i.e.~SIP/RTP, UDP, TCP or legacy applications), 
characteristics in terms of service continuity, service unavailability, end-to-end latency, security, ability to exploit simultaneously all the available NICs and to switch granularity. 
When we refer to the support provided to some communication protocols (e.g.~SIP, RTP, TCP), we mean that the considered system copes with the typical problems arising at that protocol level during a handover. 
It is thus evident that a particular scheme may provide support only to protocols working at a higher level of the network stack.

Of course, only multihoming systems (described in Section 
\ref{sec:multihoming}) may allow exploiting multiple NICs simultaneously 
(``Exploit simultaneously multiple NICs'' row). 

The row ``granularity'' reports whether or not a system works by considering i) the MN as a 
whole (N), ii) a channel where a communication flow occurs between the MN and its 
CNs (C), or iii) even when the communication management occurs on a per-packet 
basis (P). Again, systems in the same class work in a similar way. 
Thus, pure end-to-end systems work at the level of communication channels. 
Hybrid end-to-end systems and those that employ software on the home network 
operate at the level of the node (with the exception of FlowMob that work with 
channels). Systems employing specific features at the access network work at 
node or channel levels. Finally, solutions that employ some (visible or 
invisible) relay work on a per-packet basis.

It is worth noting that few systems are able to overcome the problems which may arise when NATs and firewalls are in the network. 
This is a typical feature of proxy-based approaches. 
However, the solutions that separate locators and identifiers, do not take into consideration the presence of NAT, since it is assumed that IPv6 is employed, thus NAT would not be used anymore.
Moreover, the use of identifiers enables firewalls to have access control rules 
that are based on identity, rather than address or location. However, this 
requires further work and costs to the firewall; in the case of symmetric firewalls, 
every time an end-node creates a connection using a new identifier, the 
firewall must be instructed properly. 

The unavailability interval measures the length of the interval time during 
which the MN is not able to communicate with the CN, due to the handover. This 
metric is very important when some interactive application (e.g.~VoIP) is 
executed. 
This unavailability interval includes the time needed to re-configure all the 
network entities involved in the end-to-end MN-CN communication. 
For this reason, all the solutions based on MIPv6 suffer from a high unavailability 
interval (around 1-3 seconds, typically) because they require an intense packet 
exchange between the MN, the home network and the CN to register the new 
\ac{CoA} on the \ac{HA} and to perform the return routability 
operation. 
However, if during the re-configuration of the preferred NIC, the MN exploits a different NIC, the unavailability interval decreases. 
For this reason, all the solutions employing multihoming on a per-packet basis limit the unavailability interval length. 

In addition, it is very important to note that, commonly, the unavailability time interval does not include the handover time only. When a network becomes unavailable (and thus a handover is required) there is a time interval during which the NIC appears to be functioning (at the MN), but packets are lost \cite{ngmast}. 
Then, the MN only detects that the connection with its access points has been lost after a while and eventually the handover begins. If the MN is able to detect these losses and can retransmit packets immediately, by exploiting a different, already configured and properly working NIC (as it may happen when a multihoming solution is employed), then the unavailability interval decreases. 
For that reason, the solutions that adopt some “early packet loss detector”, such as ABPS and FRHP, significantly reduce the unavailability interval.
For the sake of completeness, in the row labeled ``Unavailability Interval Due to Handover'' in Table \ref{tab2}, the symbol $\infty$ 
means that the architecture may fail to complete the handover: as concerns pure end-to-end solutions, this happens when both MN and CN change their IP addresses at the same time; in access network–based solutions, this happens when the MN enters an access network whose border gateway is not able to support the host mobility.

The ``continuity interval'' row in Table \ref{tab2} ranks the time during which 
the MN may communicate with the CN. 
In particular, it measures the distance between two consecutive unavailability 
intervals, in intermittent communication scenarios.  
This metric rewards those systems capable of switching seamlessly from the NIC in use to another NIC already working. 
Conversely, all the host mobility solutions that do not allow the simultaneous use of multiple NICs, would incur severe service interruptions; this corresponds to a low continuity of service level. 
Similarly, those architectures where handovers may fail (see the discussion of the unavailability interval above) may provide a low level of service continuity. 
Finally, the maximum continuity interval is provided by solutions based on the early packet loss detection; this is due to the fact that these approaches may recover from packet losses by switching to (and retransmitting through) a different NIC
\cite{ngmast}. 

The ``end-to-end latency'' row measures the time needed to deliver a packet from the MN 
application to the CN application. This metric is very important for interactive 
applications. 
The lowest bound of this latency corresponds to a direct communication 
between the MN and its CN, as in the case of pure and hybrid end-to-end 
host mobility solutions. 
The access network–based solutions do not introduce additional delays because 
the packets transit through an entity (the border gateway) that is already in 
the path between MN and CN. In home network-based solutions, on the other hand, the 
situation is more complex. Indeed, if the return routability operation is 
successful, then the communication between MN and CN is direct, with a 
consequent minimum latency. However, if this operation fails, then packets 
transit through the home network, hence increasing the average 
latency, depending on the distance between the CN and the home network, and 
between the home network and the MN. 
Solutions based on an external relay/proxy obviously add a delay, depending on 
the distance of the relay/proxy server. However, it is worth pointing out that 
in the real scenario there are often firewalls and NAT systems that block 
direct communication and impose the interposition of a relay. Thus, in this 
common situation there is a latency increase, except for the external 
relay–based solutions that already utilize a relay. 

The last two rows in Table \ref{tab2} report the end-to-end latency in the best 
scenario (with no firewalls) and in the most common scenario (with symmetric 
firewalls), respectively. In the most common scenario, all the solutions are 
almost equivalent in terms of end-to-end latency. 

\section{Conclusions}
\label{sec:conc}

This work presents an overview of the main architectural solutions for managing the handover process in mobile wireless communications.  
All the examined schemes have been grouped and categorized according to their main 
characteristics, the features they offer and the level of network 
stack in which they operate.  

This survey highlights that a number of issues need to be addressed when 
designing an architecture for mobile communications, even though the same issues may not be as obvious as it first appears. 
Our discussion focuses on the key issue of handover management, 
and concludes that, firstly, the best solution to this issue has 
not yet been identified. Indeed, all the schemes investigated in the literature present some advantages and drawbacks.  

Secondly, at the time of writing, the most widespread approach for building software 
architectures that support users mobility is based on the use of MIPv6. The 
solutions that adopt this approach are not deployable in current IPv4 networks.  
Moreover, these solutions cannot coexist with symmetric firewalls.

Thirdly, we have shown that many techniques are not able to cope with some problems 
due to the presence of symmetric NATs (this is a main issue for strategies 
that might be deployed on the current IPv4 Internet) and symmetric firewalls 
placed in front of two end-systems, which prevents the two end-systems to communicate. Drawing on these considerations, we can say that the approaches that solve the 
problem through the use of external relays seem to be more promising. 

Fourth, there are many popular applications and protocols that violate the protocol stack stratification, since they add information regarding the network and transport layers within application data payloads. 
A prominent example is SIP. 
These applications and protocols introduce several complications in the design of an effective mobility architecture. Indeed, to make the underlying system application/protocol-transparent, it is necessary to update the application payloads included in the transmitted packets. 
This is done to ensure that the IP address and port contained at the transport and network layers match those included within the application data. 
This task can be more easily accomplished by exploiting two proxies, a local one and an external one, working on top of the transport layer, rather than some network-layer solutions.

On the other hand, those applications and protocols that do not violate the 
protocol stratification can be easily supported by MIPv6 solutions. Note that 
this is true when only one network interface is used by the MN. As regards multihoming, the approaches may be subjected to poor performances in terms 
of introduced latencies and connection continuity intervals. 

In conclusion, the above considerations suggest that it is quite unlikely that many of these proposals will be adopted in real life scenarios, at least on a short- or medium- term basis. 
Our claim is that future proposals should deal with issues concerned with the presence of NATs and firewalls natively. 
A good solution is to resort to dedicated services such as external proxies, that can be easily distributed on the Internet. 
As a matter of fact, many applications and their related protocols include the use of external proxies, although they are usually exploited for application purposes only. 
Some examples include VoIP applications, SIP-based applications and, in certain cases, HTTP-based applications \cite{FerrettiG09}. 
We suggest that their use may be extended, in order to cater for issues concerned with mobility, as has already been proposed in the literature \cite{GhiniJSS,ghi06}.

As a concluding remark, it should be mentioned that cloud computing 
environments can be used as the infrastructure to dynamically set up (and release) 
the proxies on the server-side, in accordance with the pay-as-you-go 
principle of cloud based services \cite{cloud10}.

\end{document}